\documentclass[11pt,a4paper]{article}
\pdfoutput=1
\usepackage{jheppub}
\usepackage{latexsym}
\usepackage{pdflscape}
\usepackage{bm}
\numberwithin{equation}{section}

\allowdisplaybreaks[2]

\def\cO{\mathcal{O}}

\def\cB{\mathcal{B}}
\def\cS{\mathcal{S}}

\def\mint{\int_{-\infty}^\infty\!\cdots\!\int_{-\infty}^\infty}

\newcommand{\be}{\begin{equation}}
\newcommand{\ee}{\end{equation}}
\newcommand{\ba}{\begin{aligned}}
\newcommand{\ea}{\end{aligned}}

\DeclareMathOperator{\arctanh}{arctanh}

\DeclareMathOperator{\Li}{Li}

\def\({\left(}
\def\){\right)}

\newcommand{\pd}{\partial}

\DeclareMathOperator{\im}{Im}

\DeclareMathOperator{\Disc}{Disc}

\newcommand{\re}{{\rm e}}
\newcommand{\ri}{{\rm i}}
\newcommand{\rd}{{\rm d}}

\newcommand{\nn}{\nonumber \\}
\def\th{\theta}

\newcommand{\hf}{\frac{1}{2}}

\def\til#1{\widetilde{#1}}

\def\del{\partial}

\def\la{\lambda}

\def\rt#1{\sqrt{#1}}
\def\sitarel#1#2{\mathrel{\mathop{\kern0pt #1}\limits_{#2}}}

\preprint{DESY\ 15-079}

\title{Resummations and Non-Perturbative Corrections}

\author[a]{Yasuyuki Hatsuda}
\author[b]{and Kazumi Okuyama}

\affiliation[a]{DESY Theory Group, DESY Hamburg, \\
Notkestrasse 85, D-22603 Hamburg, Germany}
\affiliation[b]{Department of Physics, \\
Shinshu University, Matsumoto 390-8621, Japan}

\emailAdd{yasuyuki.hatsuda@desy.de} 
\emailAdd{kazumi@azusa.shinshu-u.ac.jp}

\abstract{
We consider a generalization of the Borel resummation, which turns out to be equivalent to the standard Borel resummation.
We apply it to the simplest large $N$ duality between the pure Chern-Simons theory
and the topological string on the resolved conifold, and obtain a simple integral formula 
for the free energy.
Expanding this integral representation around the large radius point at finite string coupling $g_s$,
we find that it includes not only the M-theoretic
resummation  \textit{\`a la} Gopakumar and Vafa, but also a \textit{non-perturbative correction} in $g_s$.
Remarkably, the obtained non-perturbative correction is in perfect agreement with a proposal for
membrane instanton corrections in \texttt{arXiv:1306.1734}.
Various other examples are also presented. 
}

\begin{document}

\maketitle
\renewcommand{\thefootnote}{\arabic{footnote}}
\setcounter{footnote}{0}
\setcounter{section}{0}

%%%%%%%%%%%%%%%%%%%%%%%%%%%%%%%%%%%%%%%%%%%%%%%%
%%%%%%%%%%%%%%%%%%%%%%%%%%%%%%%%%%%%%%%%%%%%%%%%
%%%%%%%%%%%%%%%%%%%%%%%%%%%%%%%%%%%%%%%%%%%%%%%%
\section{Introduction and summary}\label{sec:intro}
%%%%%%%%%%%%%%%%%%%%%%%%%%%%%%%%%%%%%%%%%%%%%%%%
%%%%%%%%%%%%%%%%%%%%%%%%%%%%%%%%%%%%%%%%%%%%%%%%
%%%%%%%%%%%%%%%%%%%%%%%%%%%%%%%%%%%%%%%%%%%%%%%%
In quantum mechanics, quantum field theories and string theory, 
perturbative expansions are usually divergent.
The standard way to resum such formal divergent series is known as Borel resummations.
In many cases, however,
the Borel resummation is not well-defined due to singularities
on the positive real axis in the Borel plane.
These singularities cause an ambiguity of the Borel resummations,
and it indicates non-perturbative corrections.
This is a standard story of the existence of the non-perturbative corrections.
Interestingly, in some (but rare) cases, perturbative expansions
are \textit{Borel summable}, but nevertheless they receive \textit{non-perturbative
corrections}.
The 't Hooft expansion of the free energy in ABJM theory \cite{ABJM} is an example for
such cases \cite{GMZ}.

In this work, we study a generalization of the Borel resummation.
An advantage of this generalization is that it is simpler than the standard Borel resummation.
In some cases, we can write down exact integral representations after the resummation.
To see this in detail, we here focus on the simplest large $N$ duality \cite{GV2}:
the duality between U($N$) pure Chern-Simons theory on a three-sphere
and the topological string on the resolved conifold.
The large $N$ expansions in these theories are well-known in all orders.
As shown in \cite{GV1, GV3}, the genus expansion of the topological string free energy
can be resummed around the large radius moduli point.
In the case of the resolved conifold, the Gopakumar-Vafa resummed formula takes the very simple form
\be
F_\text{coni}^\text{GV}(g_s,t)=\sum_{m=1}^\infty \frac{1}{m} \(2 \sin \frac{m g_s}{2} \)^{-2} \re^{-m t},
\label{eq:F-coni}
\ee
where $t$ is the K\"ahler modulus of the base $\mathbb{P}^1$ of resolved conifold.
It turns out that the original genus expansion of the free energy \eqref{eq:F-coni}
is Borel summable for $\Im(t)\not=0$.%
\footnote{Throughout this paper, we assume that the string coupling $g_s$ is real.}
Though the genus expansion is an asymptotic series,
the Gopakumar-Vafa formula \eqref{eq:F-coni} is a convergent series with respect to $\re^{-t}$.
The formula \eqref{eq:F-coni} is understood as the resummation in the M-theory regime: $t \to \infty$ with fixed finite $g_s$.
However, one easily notices that the formula \eqref{eq:F-coni} has an infinite number of poles for every rational value of $g_s/\pi$.
Thus it looks insufficient for finite $g_s$.
As we will show in this paper, we find a simple integral representation of \eqref{eq:F-coni} by applying the generalized Borel
resummation to the genus expansion:%
\footnote{For $t \in \mathbb{R}$, this resummation formula is not well-defined due to
a branch point singularity at $x=|t/g_s|$ on the positive real axis. 
One has to avoid this branch cut in the resummation.
We consider this issue in subsection~\ref{sec:non-Borel}.}
\be
F_\text{coni}^\text{resum}(g_s,t)=\frac{\Li_3(\re^{-t})}{g_s^2}
-\int_0^\infty \rd x \frac{x}{\re^{2\pi x}-1} \log (1+\re^{-2t}-2\re^{-t} \cosh g_s x ),
\label{eq:F-coni-resum}
\ee
where $\Li_n(z)$ is the polylogarithm.
One can check that the naive large $t$ expansion recovers the Gopakumar-Vafa formula \eqref{eq:F-coni},%
\footnote{To perform the integral in $\re^{-mt}$ term, one needs to assume $|g_s|<2\pi/m$. 
Once the integral is performed, it is analytically continued to $g_s \in \mathbb{R}$.
} 
but, as shown just below, the integral representation has much more information.
As we will see in section~\ref{sec:conifold}, this expression is equivalent to the standard Borel
resummation of the genus expansion. 
In \cite{HMMO}, non-perturbative corrections in topological string
on general local Calabi-Yau manifolds were proposed.
According to the formula in \cite{HMMO}, it is natural to set
\be
t = T+\pi \ri,
\label{eq:t-B}
\ee
where the second term is due to a B-field effect.
Then \eqref{eq:F-coni-resum} becomes
\begin{align}
 F_\text{coni}^\text{resum}(g_s,T+\pi\ri)=\frac{\text{Li}_3(-\re^{-T})}{g_s^2}
-\int_0^\infty \rd x\frac{x}{\re^{2\pi x}-1}\log(1+\re^{-2T}+2\re^{-T}\cosh g_sx).
\label{eq:Fconi-resumT}
\end{align}
Remarkably, this resummation formula (or \eqref{eq:F-coni-resum}) does not have any poles for finite $g_s$!
It is well-defined for any $g_s \in \mathbb{R}$.
How are the poles in \eqref{eq:F-coni} removed in \eqref{eq:Fconi-resumT}?
We observe that the integral representation \eqref{eq:Fconi-resumT} has an additional contribution to
the Gopakumar-Vafa formula \eqref{eq:F-coni}.
In fact, if one considers the ``re-expansion'' of \eqref{eq:Fconi-resumT} around the large radius point,
it turns out that \eqref{eq:Fconi-resumT} contains a non-perturbative correction
of the form $\re^{-2\pi T/g_s}$.
More explicitly, we find that the resummation \eqref{eq:Fconi-resumT} is equivalent to
\be
F_\text{coni}^\text{resum}(g_s,T+\pi\ri)=F_\text{coni}^\text{GV}(g_s,T+\pi\ri)+F_\text{coni}^\text{np}(g_s,T ),\qquad
T \to \infty,
\label{eq:F-coni-resum2}
\ee
where the first term is the original Gopakumar-Vafa contribution and the second term
is the non-perturbative contribution:
\be
F_\text{coni}^\text{np}(g_s,T)=-\sum_{\ell=1}^\infty \frac{1}{4\pi \ell^2}
\csc \( \frac{2\pi^2 \ell}{g_s} \) \left[
\frac{2\pi \ell}{g_s}T+\frac{2\pi^2 \ell}{g_s} \cot \( \frac{2\pi^2 \ell}{g_s} \)+1 \right]
\re^{-\frac{2\pi \ell T}{g_s} }.
\label{eq:F-coni-np}
\ee
This non-perturbative correction also has poles at finite $g_s$, and both poles coming
from \eqref{eq:F-coni} and \eqref{eq:F-coni-np} are precisely canceled!
As a result, the resummed free energy \eqref{eq:F-coni-resum2} is always well-defined.
This non-trivial pole cancellation is completely parallel with that in the ABJM free energy, 
originally discovered in \cite{HMO2}. 
In fact, the non-perturbative correction \eqref{eq:F-coni-np} is in perfect agreement with 
the conjecture for membrane instanton corrections in \cite{HMMO}.
The remarkable conclusion in \cite{HMMO} is that this non-perturbative correction
is constructed from the \textit{refined topological string} free energy in the so-called
\textit{Nekrasov-Shatashvili limit} \cite{NS, ACDKV}:
\be
\ba
F_\text{coni}^\text{np}(g_s,T)=\frac{1}{2\pi \ri} \frac{\pd}{\pd g_s} \left[ g_s F_\text{coni}^\text{NS}\(\frac{2\pi}{g_s},\frac{2\pi T}{g_s}\) \right],
\ea
\ee
where $F_\text{coni}^\text{NS}$ is the free energy of the refined topological string in the Nekrasov-Shatashvili limit 
(see \eqref{eq:F-coni-NS}).
As explained in \cite{HMMO}, 
we have to turn on an extra B-field in the worldsheet instanton part relative
to the membrane instanton for the pole cancellation mechanism to work.
It is interesting that this relative phase naturally appears in \eqref{eq:F-coni-resum2}.

\begin{figure}[tb]
\begin{center}
\resizebox{145mm}{!}{\includegraphics{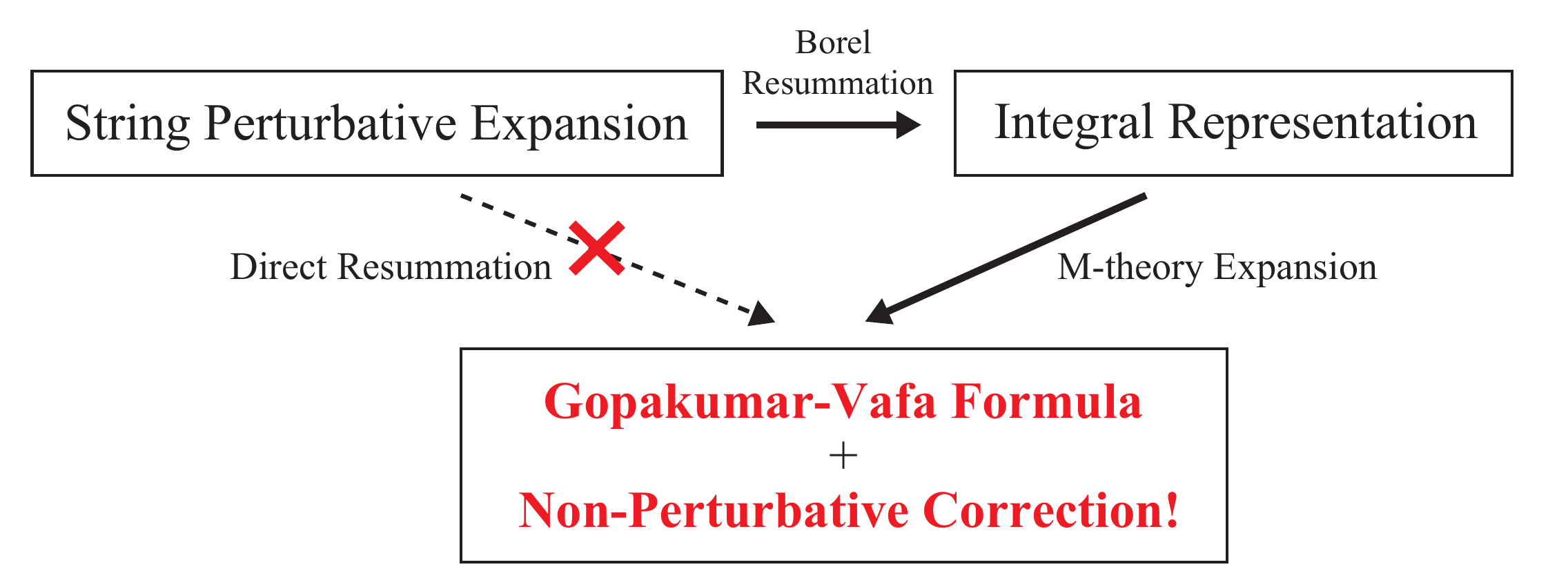}}
\end{center}
\vspace{-0.3cm}
\caption{A conceptual flow in this work. Starting with the genus expansion of the free energy,
we perform the (generalized) Borel resummation, and then obtain an integral representation.
Re-expanding this integral in the M-theory limit, we finally find the well-known Gopakumar-Vafa formula
and an additional non-perturbative correction. Note that the direct resummation of the perturbative expansion
around the large radius point leads to only the Gopakumar-Vafa formula, and the non-perturbative correction is \textit{invisible} in this way.}
  \label{fig:Flow}
\end{figure}

The emergence of the non-perturbative correction here is quite different from resurgent
trans-series expansions.
In fact, we start with the string perturbative expansion alone, and just consider its (generalized) 
Borel resummation.
Curiously, this Borel resummation leads to the additional correction to the ``perturbative'' Gopakumar-Vafa
resummation \eqref{eq:F-coni} in the M-theory regime: $T \to \infty$ with $g_s$ fixed finite.
We emphasize that we do not consider a trans-series expansion at all.
The string perturbative resummation already indicates the non-perturbative correction in the M-theoretic expansion.
We conclude that the Borel resummation of the resolved conifold free energy \eqref{eq:F-coni} 
includes the non-perturbative
correction of the form $\re^{-2\pi T/g_s}$ in the large radius limit and that it cancels the singularities from the 
Gopakumar-Vafa formula.
A flow of this structure is depicted in figure~\ref{fig:Flow}.

%%%%%%%%%%%%%
%%%%%%%%%%%%%

We find that this resummation procedure works in some other examples, but should note that
it does not always work perfectly.
In particular, it is known that the Borel resummation of the ABJM free energy does not
reproduce the exact result \cite{GMZ}.
In this case, there are additional non-perturbative corrections \cite{DMP2} to the Borel resummation, called complex instantons.
However, we stress that the Borel resummation of the ABJM free energy also removes
all the poles in the Gopakumar-Vafa formula, as was observed in \cite{GMZ}.
In section~\ref{sec:ABJM-resum}, we will see that the Borel resummation
already contains a part of the non-perturbative corrections in $g_s$.
Up to now, we do not have a resummation method to capture the full non-perturbative corrections
in the ABJM free energy  beyond the Borel sum. It would be a challenging open problem to find it out if any.

The organization of this paper is as follows.
In section~\ref{sec:Borel}, we review the procedure of the Borel resummation
and its generalization. This generalization is very useful in our analysis.
In section~\ref{sec:double-sine}, we focus on the double sine function.
The double sine function is a building block of the 3d partition function on an ellipsoid.
In particular, we see that the resummation of the vortex contribution in the double sine function
automatically contains the anti-vortex contribution as a non-perturbative correction.
In section~\ref{sec:pureCS}, we consider a resummation of the pure Chern-Simons free energy.
We find that the resummed free energy precisely reproduces the exact result.
In section~\ref{sec:conifold}, we proceed to the resummation of the resolved conifold free energy.
We show that the resummed free energy contains not only the Gopakumar-Vafa formula
but also the non-perturbative membrane instanton correction, proposed in \cite{HMMO}.
We also give a comment on the non-Borel summable case.
In section~\ref{sec:ABJM-resum}, we revisit the Borel resummation of the free energy in ABJM theory.
The 't Hooft expansion of the free energy is Borel summable, but it is known that its Borel
resummation does not reproduce the exact result.
We quantitatively evaluate the difference between the exact result and the Borel resummed one.
The finial section is devoted to conclusions of the paper.

%%%%%%%%%%%%%%%%%%%%%%%%%%%%%%%%%%%%%%%%%%%%%%%%
%%%%%%%%%%%%%%%%%%%%%%%%%%%%%%%%%%%%%%%%%%%%%%%%
%%%%%%%%%%%%%%%%%%%%%%%%%%%%%%%%%%%%%%%%%%%%%%%%

\section{Borel resummation and its generalization}\label{sec:Borel}
Let us start with a review on the Borel resummation and its simple generalization.
Such a resummation plays a crucial role in our analysis.

\subsection{Borel sum and the moment method}
In many examples of perturbative expansions in quantum field theories and string theory, we encounter
an divergent series
\begin{align}
 f(g)=\sum_{n=0}^\infty a_ng^n,
\end{align}
where the coefficients $a_n$ typically grow factorially: $a_n\sim n!$.
One useful way to make sense of such divergent series is to introduce the Borel sum
\begin{align}
 \mathcal{B}[f](\zeta)=\sum_{n=0}^\infty \frac{a_n}{n!}\zeta^n,
\label{eq:Borel-transform}
\end{align}
from which we may define the Borel resummation of the series by the Laplace transformation
\begin{align}
 \mathcal{S}_\th f=\int_0^{\infty \re^{\ri \th}} \rd \zeta\, \re^{-\zeta} \mathcal{B}[f](g\zeta).
\end{align}
Here we should choose a direction $\th$  in such a way that the 
singularity of $\mathcal{B}[f](\zeta)$ is avoided.
If $\mathcal{B}[f](\zeta)$ has no singularity along the positive real axis
on the $\zeta$-plane, then we can choose $\th=0$.

We can also consider a generalization of the Borel sum, known as the moment method, by using some moment $\mu_n$ of the weight function
$\mu(\zeta)$,
\begin{align}
 \int_0^\infty \rd \zeta\, \mu(\zeta)\zeta^n=\mu_n,
\end{align}  
and taking the sum
\begin{align}
 \mathcal{B}_\mu[f](\zeta)=\sum_{n=0}^\infty \frac{a_n}{\mu_n}\zeta^n.
\label{Bmusum}
\end{align}
From this sum $\mathcal{B}_\mu[f](\zeta)$, we can define the resummation associated with the moment $\mu_n$
\begin{align}
 \int_0^\infty \rd \zeta\, \mu(\zeta) \mathcal{B}_\mu[f](g\zeta).
\end{align}
The usual Borel sum is a special case of the moment method, corresponding to
the choice $\mu(\zeta)=\re^{-\zeta}$ and $\mu_n=n!$.
To improve the convergence property of the sum \eqref{Bmusum}, we have to choose
the moment $\mu_n$ such that $\mu_n\sim n!$.

In  string perturbation theory,
the genus-$g$ contribution 
usually grows as $(2g)!$, and hence it leads to
a divergent series.
One obvious way to deal with it is to consider the Borel sum.
However, the genus-$g$ contribution in topological string
perturbation theory sometimes contains a factor of Bernoulli number $B_{2g}$.
Therefore, it is useful to consider a moment method associated with
the integral representation of $B_{2g}$
\begin{align}
 B_{2g}=(-1)^{g-1}4g\int_0^\infty \rd x\frac{x^{2g-1}}{\re^{2\pi x}-1}.
\label{B2gint}
\end{align}
Note that the Bernoulli number grows factorially as $(2g)!$. More precisely, it behaves as
\begin{align}
 B_{2g}=
\frac{2(-1)^{g-1}(2g)!\zeta(2g)}{(2\pi)^{2g}}= \frac{2(-1)^{g-1}(2g)!}{(2\pi)^{2g}}\left(1+\frac{1}{2^{2g}}+\frac{1}{3^{2g}}+\cdots\right),\quad
g \to \infty.
\end{align}
In this paper, we will call the moment method associated with the Bernoulli number as
{\it the Bernoulli moment method}.

We are interested in the resummation of the series of the form
\begin{align}
 \varphi(z)=\sum_{g=2}^\infty z^{2g-2} F_g.
\end{align}
By the Bernoulli moment method,
the resummation of this series is given by
\begin{align}
 \varphi^\text{resum}(z)=\int_0^\infty \rd x\frac{x}{\re^{2\pi x}-1}\sum_{g=2}^\infty \frac{(-1)^{g-1}4g}{B_{2g}}(xz)^{2g-2}F_g.
\label{Bmethod}
\end{align}
\subsection{Preliminary examples: gamma function and Barnes $G$-function}
As simple examples of the Bernoulli moment method,
let us consider the resummation of the asymptotic expansion
of the gamma function $\Gamma(z)$ and the Barnes $G$-function $G_2(z)$. 
The resummation of  the Barnes $G$-function
is particularly important for the Gaussian matrix model at large $N$.

The logarithm of the gamma function has the following asymptotic expansion in the limit $|z|\to \infty$
\begin{align}
 \log \Gamma(z)=z\log z-z-\hf\log\frac{z}{2\pi}+\sum_{g=1}^\infty \frac{B_{2g}}{2g(2g-1)z^{2g-1}}.
\end{align}
This is a divergent series due to the $(2g)!$ growth of the Bernoulli number
$B_{2g}$. As we explained in the previous subsection,
this series can be resummed by the Bernoulli moment method
\eqref{Bmethod} as
\begin{align}
\log \Gamma(z)=z\log z-z-\hf\log\frac{z}{2\pi}+2 \int_0^\infty \rd x\frac{\arctan(x/z)}
{\re^{2\pi x}-1}.
\end{align}
This is known as the Binet's second formula, which is valid for $\Re (z)>0$.
Interestingly, in this case the resummation of the
asymptotic series gives the {\it exact} answer.
On the other hand, the usual Borel resummation leads to the so-called Binet's first formula
\begin{align}
 \log \Gamma(z)=z\log z-z-\hf\log\frac{z}{2\pi}+
\int_0^\infty \frac{\rd x}{x}\left[\frac{1}{\re^x-1}-\frac{1}{x}+\hf\right]\re^{-zx}.
\end{align}

Next, let us consider the asymptotic expansion of the Barnes $G$-function
in the limit $|z|\to \infty$
\begin{align}
 \log G_2(z+1)=\hf z^2\log z-\frac{3}{4}z^2+\hf z\log(2\pi)-\frac{1}{12}\log z+\zeta'(-1)
+\sum_{g=2}^\infty \frac{B_{2g}}{2g(2g-2)z^{2g-2}}.
\label{Barnesexp}
\end{align}
Again, this can be resummed by using \eqref{Bmethod}
\begin{align}
 \log G_2(z+1)= \hf z^2\log z-\frac{3}{4}z^2+\hf z\log(2\pi)
+\int_0^\infty \rd x\frac{x}{\re^{2\pi x}-1}\log\left(\frac{x^2}{x^2+z^2}\right).
\label{Barnesresum}
\end{align}
Here we have used the formula
\begin{align}
 \int_0^\infty \rd x\frac{x}{\re^{2\pi x}-1}=\frac{1}{24},\qquad
\int_0^\infty \rd x\frac{x\log x}{\re^{2\pi x}-1}=\hf\zeta'(-1),
\label{genus-one-int}
\end{align}
to absorb the two terms $-\frac{1}{12}\log z+\zeta'(-1)$ in \eqref{Barnesexp} into the $x$-integral in \eqref{Barnesresum}.
We have checked numerically that \eqref{Barnesresum}
correctly reproduces the value of $\log G_2(z+1)$ for $\Re(z)>0$.

\section{Double sine function and vortex/anti-vortex partition function}\label{sec:double-sine}
The double sine function $s_b(z)$ appears in the integrand
of partition functions of three-dimensional $\mathcal{N}=2$ theories on
the ellipsoid $S_b^3$ \cite{Hama:2011ea}
\begin{align}
 b^2|z_1|^2+b^{-2}|z_2|^2=1,
\end{align} 
and $s_b(z)$ itself can  also be interpreted as the partition function of free chiral multiplet
on $S_b^3$.
In the limit $b\to0$, the ellipsoid becomes
$\mathbb{R}^2\times S^1$, and the partition function counts
the contributions of vortices on $\mathbb{R}^2$.
In the other limit  $1/b\to0$, the ellipsoid degenerates to
another copy of $\mathbb{R}^2\times S^1$. In this case the partition function counts
the contributions of anti-vortices on $\mathbb{R}^2$.
Note that in the small $b$ expansion vortices appear  perturbatively, while  
anti-vortices are non-perturbative in $b$.  
As shown in \cite{Pasquetti},
the ellipsoid partition function is
factorized into partition functions of vortices and anti-vortices, and
the ellipsoid partition function can be regarded as a non-perturbative completion of 
the vortex partition function.

In this section, we consider the
resummation of the perturbative expansion of the double sine function $s_b(z)$,
and see how the non-perturbative correction appears. 
The logarithm of double sine function $s_b(z)$ is expanded as (see e.g.
appendix A in \cite{Pasquetti})
\begin{align}
 \ri \log s_b(z)=\frac{\pi z^2}{2}+
\sum_{\ell=1}^\infty\frac{(-1)^{\ell}}{\ell}
\left[\frac{\re^{2\pi\ell bz}}{2\sin\pi\ell b^2}+\frac{\re^{2\pi\ell z/b}}{2\sin\pi\ell/ b^2}\right].
\label{logSb}
\end{align}
As discussed in \cite{Pasquetti}, the last two terms can be interpreted
as the contribution of vortices and anti-vortices,
which we will call the
 ``free energy'' of  vortices/anti-vortices
\be
\ba
F_\text{vor}(b,v)&=
\sum_{\ell=1}^\infty\frac{(-1)^{\ell}\re^{-\ell bv}}{2\ell\sin\pi\ell b^2},\\
F_{\overline{\text{vor}}}(b,v)&=\sum_{\ell=1}^\infty\frac{(-1)^{\ell}\re^{-\ell v/b}}
{2\ell\sin\pi\ell /b^2}=
F_\text{vor}(b^{-1},v).
\ea
\label{eq:Fvor}
\ee
Here we have set $v=-2\pi z$.
We will call the term proportional to $\re^{-\ell bv}$ (or $\re^{-\ell v/b}$) as the $\ell$-vortex
contribution (or $\ell$-anti-vortex contribution), respectively.
Note that as in the Gopakumar-Vafa formula \eqref{eq:F-coni}, the vortex/anti-vortex free energy 
\eqref{eq:Fvor} has poles for rational $b^2$, but these poles are totally canceled in 
the double sine function \eqref{logSb}.
As a consequence, the double sine function is well-defined for any $b$.

\begin{figure}[tb]
\begin{center}
\begin{tabular}{ccc}
\resizebox{45mm}{!}{\includegraphics{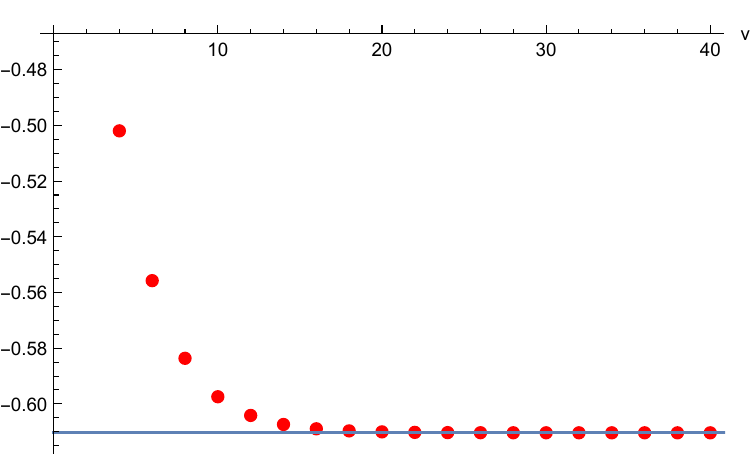}}
\hspace{2mm}
&
\resizebox{45mm}{!}{\includegraphics{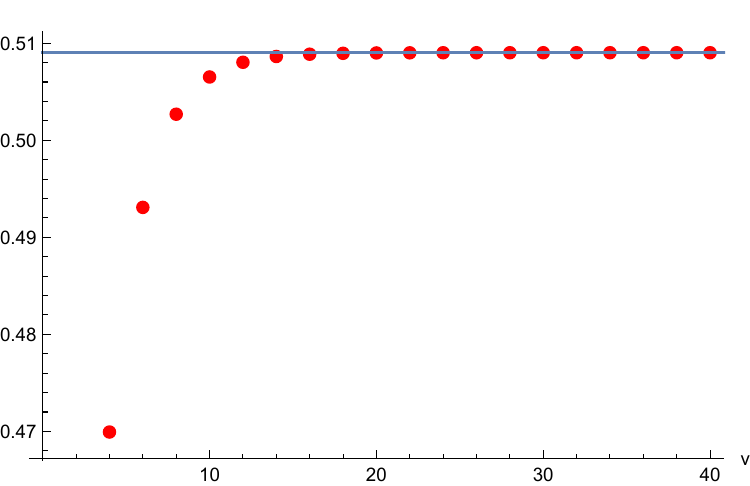}}
\hspace{2mm}
&
\resizebox{45mm}{!}{\includegraphics{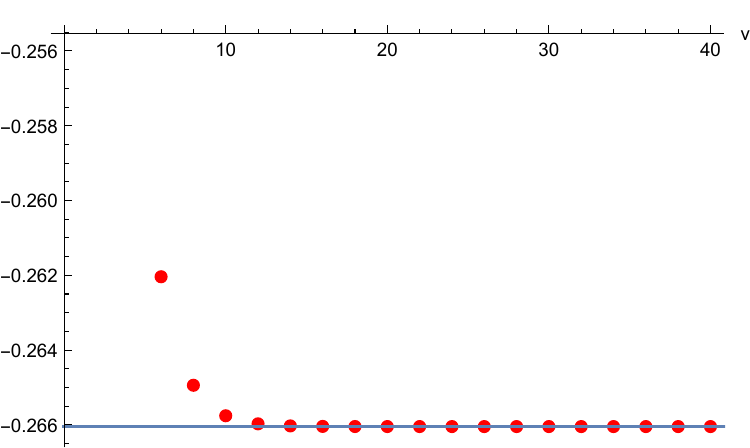}}
\\
(a) 1-anti-vortex &(b)  1-vortex
& (c) 2-anti-vortex 
\end{tabular}
\end{center}
\caption{We show the plots of $F_\text{vor}^\text{resum}(b,v)$ as a function of $v$
for $b=6/5$. The solid lines represent the expected
vortex or anti-vortex coefficients, while the
red dots represent the numerical values of the integral \eqref{eq:Fvor-resum}.
More precisely, the red dots in each figure represent: 
(a) $F_\text{vor}^\text{resum}(b,v)\re^{v/b}$, 
(b) $\big[F_\text{vor}^\text{resum}(b,v)-F_{\overline{\text{vor}}}^{(1)}(b,v)\big]\re^{bv}$,
(c) $\big[F_\text{vor}^\text{resum}(b,v)-F_{\overline{\text{vor}}}^{(1)}(b,v)-F_{\text{vor}}^{(1)}(b,v)\big]\re^{2v/b}$.
Here $F_{\text{vor}}^{(\ell)}$ (or $F_{\overline{\text{vor}}}^{(\ell)}(b,v)$)
denotes the $\ell$-vortex (or $\ell$-anti-vortex) contribution, respectively.}
  \label{fig:Fvor-resum}
\end{figure}

By using the formula
\begin{align}
 \frac{1}{2\sin\frac{z}{2}}=\sum_{g=0}^\infty \frac{(-1)^gB_{2g}(1/2)}{(2g)!}z^{2g-1},
\end{align}
where $B_{2g}(1/2)$ is the Bernoulli polynomial $B_{2g}(z)$ at $z=1/2$,
the small $b$ expansion of the vortex free energy
$F_\text{vor}(b,v)$ with fixed $bv$ is found to be
\begin{align}
 F_\text{vor}(b,v)=
\sum_{g=0}^\infty \frac{(-1)^g B_{2g}(1/2)}{(2g)!}
\text{Li}_{2-2g}(-\re^{-bv})(2\pi b^2)^{2g-1}.
\label{Fvor-smallb}
\end{align}
This small $b$ expansion
\eqref{Fvor-smallb} is an asymptotic series, but
we can resum  this series by the moment method associated with
the integral representation of  $B_{2g}(1/2)$
\begin{align}
 B_{2g}(1/2)=(-1)^{g}4g\int_0^\infty \rd x\frac{x^{2g-1}}{\re^{2\pi x}+1},\qquad(g\geq1),
\end{align}
and we find
\begin{align}
  F_\text{vor}^\text{resum}(b,v)=\frac{\text{Li}_2(-\re^{-bv})}{2\pi b^2}
+\int_0^\infty \frac{\rd x}{\re^{2\pi x}+1}
\log\left(\frac{1+\re^{-bv-2\pi b^2x}}{1+\re^{-bv+2\pi b^2x}}\right).
\label{eq:Fvor-resum}
\end{align}
Remarkably, this resummed vortex free energy contains not
only the vortex contribution but also the anti-vortex contribution!%
\footnote{The similar analysis was performed in the published version of \cite{Pasquetti}.
There, an analytic continuation of the Borel resummation from $b \in \mathbb{R}$ (Borel summable) to $b \in \ri \mathbb{R}$ 
(non-Borel summable) was discussed.
It was observed if one picks up all the pole contributions in the first quadrant, then
they produce the anti-vortex contribution correctly.
Here, we emphasize that the resummation \eqref{eq:Fvor-resum} already includes the anti-vortex part even when
the small $b$ expansion \eqref{Fvor-smallb} is \textit{Borel summable}. 
This perspective looks different from the one in \cite{Pasquetti}. 
}
We emphasize that we started with only the vortex free energy.
We did not use any information on the anti-vortex contribution.
The resummation \eqref{eq:Fvor-resum} naturally contains the anti-vortex contribution
as the non-perturbative correction:
\begin{align}
 F_\text{vor}^\text{resum}(b,v)=F_\text{vor}(b,v)+F_{\overline{\text{vor}}}(b,v).
\end{align}
To see it, let us consider the case of $b=6/5$, for example.
In this case the exponential factors of each vortex/anti-vortex
contributions have the following ordering of magnitudes
\begin{align}
 \re^{-\frac{v}{b}} > \re^{-bv} > \re^{-\frac{2v}{b}} > \cdots,\qquad (v\to\infty).
\label{order-vor}
\end{align}
As we can see in figure \ref{fig:Fvor-resum},
vortices
and anti-vortices
indeed appear in $F_\text{vor}^\text{resum}(b,v)$ in the expected ordering
\eqref{order-vor}
with the correct coefficients.
We have also checked numerically for some other values of $b$
that $F_\text{vor}^\text{resum}(b,v)$
correctly includes both vortices
and anti-vortices.
We have also checked numerically that
resummed free energy $F_\text{vor}^\text{resum}(b,v)$ \eqref{eq:Fvor-resum}
is invariant under $b\to b^{-1}$
\begin{align}
  F_\text{vor}^\text{resum}(b^{-1},v)=F_\text{vor}^\text{resum}(b,v),
\end{align}
although this symmetry is not manifest in  \eqref{eq:Fvor-resum}.
An important lesson of this example is that
some part of non-perturbative effects is already included in the
resummation of the perturbative series, and in a lucky situation, 
like in this case of vortex free energy, the
resummation of the perturbative series gives the complete non-perturbative answer.

\section{Large $N$ expansions in pure Chern-Simons theory}\label{sec:pureCS}
%%%%%%%%%%%%%%%%%%%%%%%%%%%%%%%%%%%%%%%%%%%%%%%%
%%%%%%%%%%%%%%%%%%%%%%%%%%%%%%%%%%%%%%%%%%%%%%%%
%%%%%%%%%%%%%%%%%%%%%%%%%%%%%%%%%%%%%%%%%%%%%%%%
In this section, we will consider our resummation procedure for the pure Chern-Simons theory.
We first see the detailed computation for the U($N$) gauge group,
then generalize it to the SO/Sp gauge groups.

Let us start with the exact partition function on
$S^3$ of the U($N$) pure Chern-Simons theory
at level $k>0$ \cite{Witten:1988hf}:
\be
Z_\text{CS}(N, g_s)=\left(\frac{g_s}{2\pi}\right)^{\frac{N}{2}} \prod_{j=1}^{N-1} 
\left(2\sin\frac{g_s j}{2}\right)^{N-j},
\label{eq:ZCS-exact}
\ee
where $g_s$ denotes the string coupling in the dual topological string
\begin{align}
 g_s=\frac{2\pi}{k+N}.
\end{align}
In the 't Hooft limit 
\begin{align}
 N\to\infty ,\quad \text{with}~~t=\ri g_sN~~\text{fixed},
\end{align} 
the free energy of pure CS theory has the following genus expansion:
\be
F_\text{CS}=\log Z_\text{CS}=\sum_{g=0}^\infty g_s^{2g-2} F_g(t),
\label{eq:FCS}
\ee
where the genus $g$ free energy is explicitly written in a closed form
\be
\ba
F_0(t)&=\frac{1}{2}(\Li_3(\re^{t})+\Li_3(\re^{-t}))+c_0, \\
F_1(t)&=\frac{1}{24}(\Li_1(\re^{t})+\Li_1(\re^{-t}))+c_1,\\
F_{g \geq 2}(t)&=
\frac{(-1)^{g-1}B_{2g}}{2g(2g-2)!}\Li_{3-2g}(\re^{-t})+c_g.
\ea
\ee
The $t$-independent parts $c_g$'s are known as the constant map contributions,
whose explicit forms are given by
\begin{align}
c_0=-\zeta(3),
\qquad
c_1= \zeta'(-1)+\frac{1}{12} \log g_s ,\qquad
c_{g\geq2}=\frac{(-1)^{g-1}B_{2g}B_{2g-2}}{2g(2g-2)(2g-2)!}.
\label{Fconst}
\end{align}
Since, for physical cases, both $N$ and $k$ are integers,
the 't Hooft parameter $t$ must be purely imaginary.
One can confirm that the expansion \eqref{eq:FCS} indeed give a good approximation\footnote{%
To be precise, the expansion \eqref{eq:FCS} is asymptotic for finite $t$, thus
one need to truncate it up to certain appropriate order.
} of $F_\text{CS}$ for 
small $g_s$ with $t \in \ri \mathbb{R}$.
Note that $\Li_{3-2g}(\re^{t})=\Li_{3-2g}(\re^{-t})$ for $g \geq 2$ but this does not hold for $g=0,1$.

\subsection{Resumming the U($N$) pure CS free energy}
We now consider the resummation of $F_\text{CS}$ for $t \in \ri \mathbb{R}$. 
We first notice that the constant map contribution
\eqref{Fconst} can be resummed in all genera
\cite{KEK, HO}
\be
F_\text{const}(g_s)=\frac{c_0}{g_s^2}+c_1
+\sum_{g=2}^\infty g_s^{2g-2} c_g
=\hf A_\text{c}(4\pi/g_s),
\label{constsum}
\ee
where $A_\text{c}(k)$ is a function appearing in the ABJM Fermi-gas.
In \cite{HO}, an explicit integral representation
of $A_\text{c}(k)$ was found to be
\be
A_\text{c}(k)=\frac{2\zeta(3)}{\pi^2 k}\(1-\frac{k^3}{16}\)
+\frac{k^2}{\pi^2} \int_0^\infty \rd x \frac{x}{\re^{k x}-1}\log(1-\re^{-2x}).
\label{eq:A-int}
\ee
The factor $1/2$ in \eqref{constsum} can be understood from the fact that
the $F_\text{const}(g_s)$ is proportional to $\chi/2$, where $\chi$ is the (effective)
Euler number of the target space, and $\chi=2$ for the resolved conifold while
$\chi=4$ for the local $\mathbb{P}^1\times \mathbb{P}^1$.

We can re-derive this expression
\eqref{eq:A-int} by using the Bernoulli moment method \eqref{Bmethod}.
Using the formula,
\begin{align}
\sum_{g=2}^\infty \frac{B_{2g-2}}{(2g-2)(2g-2)!}
z^{2g-2}=\log\left(\frac{2}{z}\sinh\frac{z}{2}\right),
\end{align}
the constant map contributions can be resummed as
\begin{align}
F_\text{const}(g_s)=-\frac{\zeta(3)}{g_s^2}+
2\int_0^\infty \rd x \frac{x}{\re^{2\pi x}-1}\log\Big(2\sinh\frac{g_sx}{2}\Big).
\label{Fconstresum}
\end{align}
From \eqref{genus-one-int}, one can see that \eqref{Fconstresum}
indeed agrees with \eqref{eq:A-int} for $g_s=4\pi/k$.

\begin{figure}[tb]
\begin{center}
%\hspace{-3mm}
\resizebox{70mm}{!}{\includegraphics{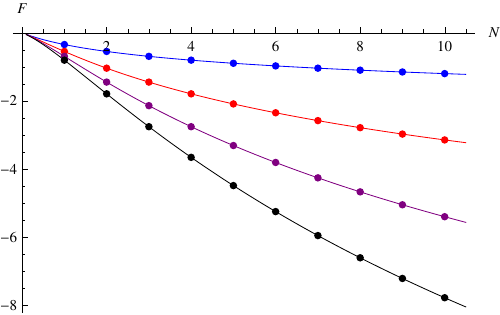}}
\end{center}
  \caption{The comparison of the resummed free energy \eqref{eq:FCS-resum} and the exact one.
 The solid curves represent the resummed free energies while the dots represent the exact values.
 The blue, red, purple and black curves (dots) correspond to $k=1,2,3,4$, respectively.}
  \label{fig:FCS-graph}
\end{figure}

Next consider the resummation of the $t$-dependent part
$\til{F}_g(t)=F_g(t)-c_g$.
Again, since $\til{F}_g$ is proportional to the 
Bernoulli number, we can use the Bernoulli moment method
\eqref{Bmethod}.
We find
\begin{align}
 \til{F}_\text{CS}^{\text{resum}}=
\frac{\til{F}_0(t)}{g_s^2}-
\int_0^\infty \rd x \frac{x}{\re^{2\pi x}-1}\log\Big(4\sinh^2\frac{g_sx}{2}-4\sinh^2\frac{t}{2}\Big).
\label{Ftilresum}
\end{align}
Finally, adding \eqref{Fconstresum} and
\eqref{Ftilresum},
the 
resummation of the pure  CS free energy is given by
\be
F_\text{CS}^\text{resum}(g_s,t)=\frac{F_0(t)}{g_s^2}+\int_0^\infty \rd x \frac{x}{\re^{2\pi x}-1}
\log \( \frac{\sinh^2 \frac{g_s x}{2} }{\sinh^2 \frac{g_s x}{2}-\sinh^2 \frac{t}{2}} \).
\label{eq:FCS-resum}
\ee
In figure~\ref{fig:FCS-graph}, we show the resummed free energy \eqref{eq:FCS-resum} and
the exact free energy computed from \eqref{eq:ZCS-exact} as functions of $N$
at fixed $k=1,2,3,4$. 
We find that the resummed free energy \eqref{eq:FCS-resum}
reproduces the exact values of pure CS free energy at finite $N$.%
\footnote{Setting the option {\tt WorkingPrecision$\to$100} for {\tt NIntegrate} in {\tt Mathematica},
we found a 100-digit agreement.
This strongly suggests that the agreement 
between $F_\text{CS}^\text{resum}$ and $\log Z_\text{CS}$ at finite $N$ is {\it exact}.}
Moreover, the resummed free energy \eqref{eq:FCS-resum}
gives a natural extension to non-integer $N,k$.%
\footnote{Another way of the analytic continuation to non-integer $N$ 
is by applying the Euler-Maclaurin formula
to the exact result \eqref{eq:ZCS-exact}. We observed that both analytic continuations lead to
the same result for non-integer $N$, but the integral \eqref{eq:FCS-resum} is much simpler.}

As we mentioned above, for the physical pure CS theory,
the 't Hooft parameter $t$ is purely imaginary, hence it is natural to set $t=2\pi \ri\la$,
where $\la$ is given by
\begin{align}
 \la=\frac{g_sN}{2\pi}=\frac{N}{k+N}.
\end{align}
As a consequence of the level-rank duality,
the free energy of pure CS theory is invariant under
$\la\to1-\la$ with $g_s$ fixed (see e.g. \cite{Kapustin:2010mh} for a proof). 
One can easily see that
the resummed free energy \eqref{eq:FCS-resum}
is indeed invariant under $\la\to1-\la$.

It is interesting to observe that in the limit $g_s\to0$
the last term in \eqref{eq:FCS-resum}
reduces to the integral appearing in the resummation of the Barnes function $\log G_2(N+1)$
in \eqref{Barnesresum}.
This is expected since the partition function of pure CS theory
is proportional to the $q$-Barnes function $G_q(N+1)$ with $q=\re^{\ri g_s}$.

\subsection{Pure Chern-Simons with SO/Sp gauge groups}
One can also consider 
the resummation of the
genus expansion of the free energy of pure CS theory
with
SO($N$) or Sp($N$) gauge groups.
As shown in \cite{Sinha:2000ap},
they are dual to the topological string on the resolved conifold, 
modded out by a $\mathbb{Z}_2$ orientifold projection.
The exact $S^3$ partition function has a similar form as the U($N$) case
\begin{align}
 Z_\text{CS}&=\left(\frac{g_s}{2\pi}\right)^{\frac{r}{2}}\prod_{j}
\Big(2\sin\frac{g_sj}{2}\Big)^{f(j)},
\label{ZCSsosp}
\end{align}
where $r$
denotes the rank of gauge group, 
and $f(j)$ is the multiplicity, whose explicit form can be found in \cite{Sinha:2000ap}.
In  \eqref{ZCSsosp}, 
we have ignored the overall normalization factor $|P/Q|^{-1/2}$, coming from 
the difference of the weight lattice $P$ and the root lattice $Q$.
The string coupling $g_s$ for the SO($N$) and the Sp($N$) cases are given by
\begin{align}
 g_s=
\begin{cases}
 \displaystyle\frac{2\pi}{k+N-2},\quad &\text{for}~~ \text{SO}(N),\\
&\\
\displaystyle\frac{2\pi}{k+N/2+1},\quad &\text{for}~~ \text{Sp}(N).
\end{cases}
\end{align}

The large $N$ expansion of 
the free energy of SO/Sp pure Chern-Simons theory takes the form \cite{Sinha:2000ap}
\begin{align}
 F_{\text{SO/Sp}}=\hf F_{\text{U}}\pm F_\text{non-ori},
\label{sosp}
\end{align}
where $F_{\text{U}}$ is the 
free energy of pure CS theory with U($N$) gauge group, and
$F_\text{non-ori}$ is the contribution
coming from the non-orientable worldsheets due to
orientifold projection.

\begin{figure}[tb]
\begin{center}
\begin{tabular}{ccc}
%\hspace{-3mm}
\resizebox{45mm}{!}{\includegraphics{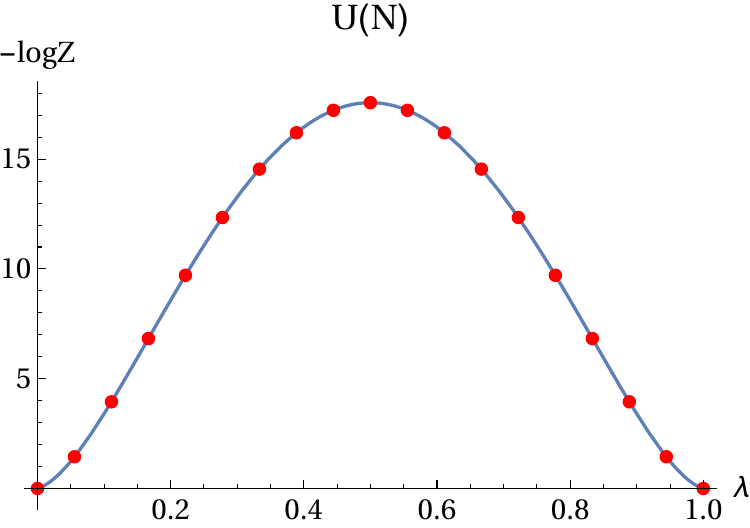}}
\hspace{2mm}
&
\resizebox{45mm}{!}{\includegraphics{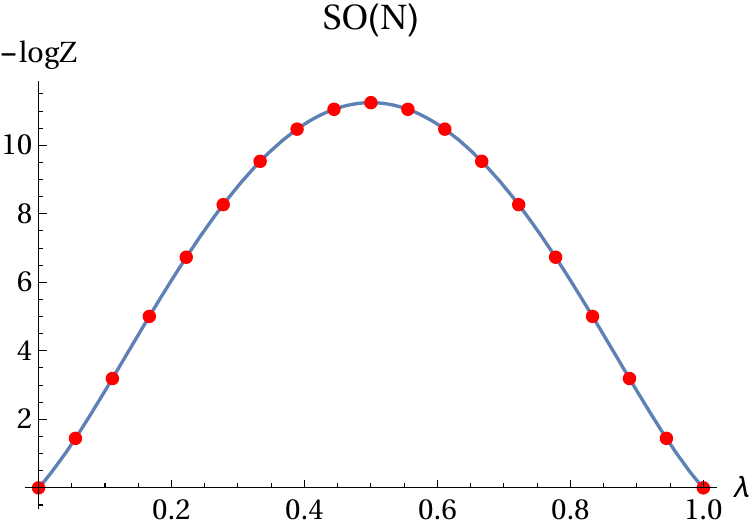}}
\hspace{2mm}
&
\resizebox{45mm}{!}{\includegraphics{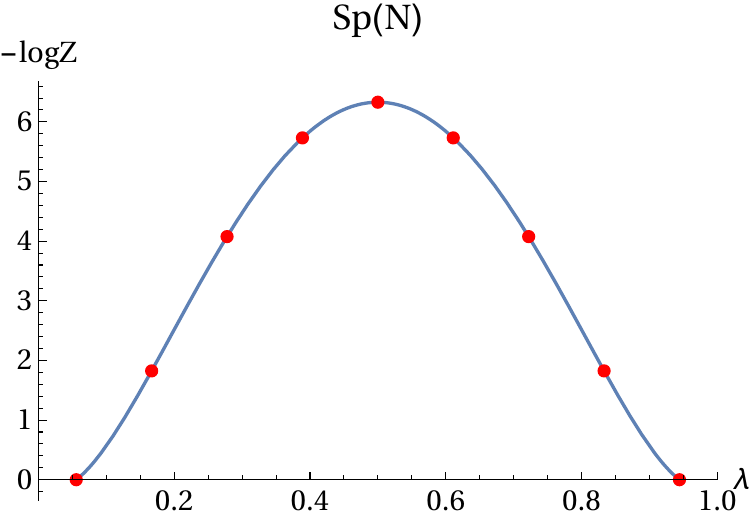}}
%\vspace{-5mm}
\end{tabular}
\end{center}
  \caption{We show the minus of the free energy $-\log Z$ for 
pure CS theories with U$(N)$ (left), SO$(N)$ (middle), and Sp($N$) (right)
gauge groups at $g_s=\pi/9$ in the range $0\leq\la\leq1$.
The solid curves represent the resummed free energy, while the red dots
show the exact values $-\log Z_\text{CS}$ at integer $N$.}
  \label{fig:SUSOSP-pureCS}
\end{figure}

Here we focus on the physical theory, corresponding to the pure
imaginary K\"{a}hler parameter $t=2\pi\ri \la$
with the 't Hooft coupling $\la$ given by
\begin{align}
 \la=\frac{(N\mp1)g_s}{2\pi},\qquad \text{for}~~\text{SO}(N)/\text{Sp}(N).
\end{align}
The resummed free energy of U($N$) CS theory is given  by
\eqref{eq:FCS-resum} with the replacement $t=2\pi\ri\la$
\begin{align}
 F_{\text {U}}^\text{resum}=
\frac{\text{Re}[\text{Li}_3(\re^{-2\pi \ri\la})]-\zeta(3)}{g_s^2}
+\int_0^\infty \rd x\frac{x}{\re^{2\pi x}-1}\log\left(\frac{\sinh^2\frac{g_sx}{2}}{\sinh^2\frac{ g_sx}{2}+\sin^2\pi \la}\right).
\label{Fsu-resum}
\end{align}
On the other hand, $F_\text{non-ori}$
is given by the following Gopakumar-Vafa type formula \cite{Sinha:2000ap}
\begin{align}
 F_\text{non-ori}=-\sum_{\ell=\text{odd}}\frac{\sin\pi \ell\la}{2\ell\sin\frac{g_s\ell}{2}}.
\label{GVnon-ori}
\end{align}
However, this is just a formal expression, and hence we have to 
perform a resummation of this series to make sense of it at finite
$g_s$. In a similar manner as the vortex free energy in the previous section,
we can resum this series as
\begin{align}
F_\text{non-ori}^\text{resum}
=\frac{\text{Im}\big[\text{Li}_2(\re^{-\ri\pi \la})-\text{Li}_2(-\re^{-\ri\pi \la})\big]}{2g_s}
+\int_0^\infty \frac{\rd x}{\re^{2\pi x}+1}\arctan\left(\frac{\sin \pi \la}{\sinh g_sx}\right).
\label{Fnon-resum} 
\end{align}
Finally,  the resummed free energy of
SO($N$)/Sp($N$)
pure CS theory is given by
$\hf F_\text{U}^\text{resum}\pm F_\text{non-ori}^\text{resum}$ 
with \eqref{Fsu-resum} and \eqref{Fnon-resum}.
We have checked numerically that resummed free energy
correctly reproduces the exact partition function
\eqref{ZCSsosp}
at finite $N$.
Again, the resummed free energy
defines a natural extension to non-integer $N,k$.
One can see that  $F_\text{U}^\text{resum}$ in 
\eqref{Fsu-resum} and $F_\text{non-ori}^\text{resum}$ in \eqref{Fnon-resum}
are invariant under $\la\to1-\la$ with fixed $g_s$. Also note that, as observed in \cite{Sinha:2000ap},
the
non-orientable part $F_\text{non-ori}^\text{resum}$ changes sign by the shift $\la\to\la+1$.
In figure \ref{fig:SUSOSP-pureCS}, we show the plot of
the (minus of) resummed free energy for U$(N)$, SO$(N)$, and Sp($N$) pure CS theories
at $g_s=\pi/9$, together with the exact free energy at various integer $N$
in the range $0\leq\la\leq1$. 
One can clearly see an excellent agreement for all three cases.

\section{Resummation of the resolved conifold free energy}\label{sec:conifold}
Let us proceed to the topological string free energy on the resolved conifold.
It is well-known that the large $N$ expansion of the pure Chern-Simons theory is equivalent to the genus expansion 
of the topological string free energy on the resolved conifold via the large $N$ duality \cite{GV2}:
\be
F_\text{CS}(g_s,t)=F_\text{coni}(g_s,t).
\label{eq:largeNduality}
\ee
The 't Hooft coupling $t$ just corresponds to the K\"ahler modulus in the topological string.
As mentioned before, for the physical case of the Chern-Simons theory, $t$ must be
purely imaginary, but one can analytically continue it to complex values.

\subsection{Resumming the resolved conifold free energy}
When $\Re(t)\gg1$,
it is useful to rewrite the free energy as the Gopakumar-Vafa form \eqref{eq:F-coni}.
It is easy to see that \eqref{eq:F-coni} can be
expanded in $g_s$ as
\begin{align}
 F_\text{coni}^\text{GV}(g_s,t)=\frac{\text{Li}_3(\re^{-t})}{g_s^2}+
\frac{1}{12}\text{Li}_1(\re^{-t})+\sum_{g=2}^\infty g_s^{2g-2}\til{F}_g(t),
\label{eq:FGVexp}
\end{align}
where
\be
\til{F}_{g \geq 2}(t)=
\frac{(-1)^{g-1}B_{2g}}{2g(2g-2)!}\Li_{3-2g}(\re^{-t}).
\ee 
Of course, the same result is obtained from the large $N$ expansion of the U($N$) pure CS free energy \eqref{eq:FCS}.
Using the relations
\be
\ba
\frac{1}{2} \left[ \Li_3(\re^{t})+\Li_3(\re^{-t}) \right]&=-\frac{t^3}{12}
+\frac{\pi \ri}{4}t^2+\frac{\pi^2}{6}t+\Li_3(\re^{-t}) ,\\
\frac{1}{24}\left[\text{Li}_1(\re^{t})+\text{Li}_1(\re^{-t})\right]
&=-\frac{t-\pi \ri}{24}+\frac{1}{12}\text{Li}_1(\re^{-t}),
\ea
\ee
one finds that the exponentially suppressed part of \eqref{eq:FCS} in $\Re(t) \to \infty$
indeed reproduces \eqref{eq:FGVexp}.
One can also show that \eqref{eq:FGVexp} can be resummed
in the same manner as \eqref{Ftilresum}.
The final result is given by \eqref{eq:F-coni-resum}.
When $t$ is real, the resummation formula \eqref{eq:F-coni-resum} is not well-defined because
the integrand has a branch point singularity at $x=|t/g_s|$ on the positive real axis. 
In this case, one has to avoid the branch cut in the integral.
We will consider this case in the next subsection.

As discussed in the introduction, here we set $t=T+\pi\ri$
and consider the resummed free energy $F_\text{coni}^\text{resum}(g_s,T+\pi\ri)$ in \eqref{eq:Fconi-resumT}. Very surprisingly
we find that $F_\text{coni}^\text{resum}(g_s,T+\pi\ri)$ in \eqref{eq:Fconi-resumT} contains the 
contributions of membrane instantons,\footnote{%
At the level of the integrand in \eqref{eq:Fconi-resumT}, there is no non-perturbative correction in $g_s$.
The non-perturbative correction of the form $\re^{-2\pi T/g_s}$ appears as a consequence of the integral.
This mechanism is quite similar to a recent proposal in the ABJM Fermi-gas system \cite{Hatsuda}.
}
which is precisely given by
the refined topological string in the Nekrasov-Shatashvili limit as proposed in \cite{HMMO}
(see also \cite{Hatsuda})
\begin{align}
F_\text{coni}^\text{resum}(g_s,T+\pi\ri)
=F_\text{coni}^\text{GV}(g_s,T+\pi \ri)
+\frac{1}{2\pi \ri} \frac{\pd}{\pd g_s} \left[ g_s F_\text{coni}^\text{NS}\(\frac{2\pi}{g_s},\frac{2\pi T}{g_s}\) \right],
\label{FM2FWSconi}
\end{align}
where the free energy in the Nekrasov-Shatashvili limit is given by
\be
F_\text{coni}^\text{NS}(g_s,t)=\frac{1}{2\ri} \sum_{\ell=1} \frac{\re^{-\ell t}}{\ell^2 \sin (\ell g_s)}.
\label{eq:F-coni-NS}
\ee
%%%%%%%%%%%%%%%%%%%%%%%%%%%%%%%%%%%%%%%%%%%%%%%%%%%%
%%%%%%%%%%%%%%%%%%%%%%%%%%%%%%%%%%%%%%%%%%%%%%%%%%%%
\begin{figure}[tb]
\begin{center}
\begin{tabular}{cc}
%\hspace{-3mm}
\resizebox{50mm}{!}{\includegraphics{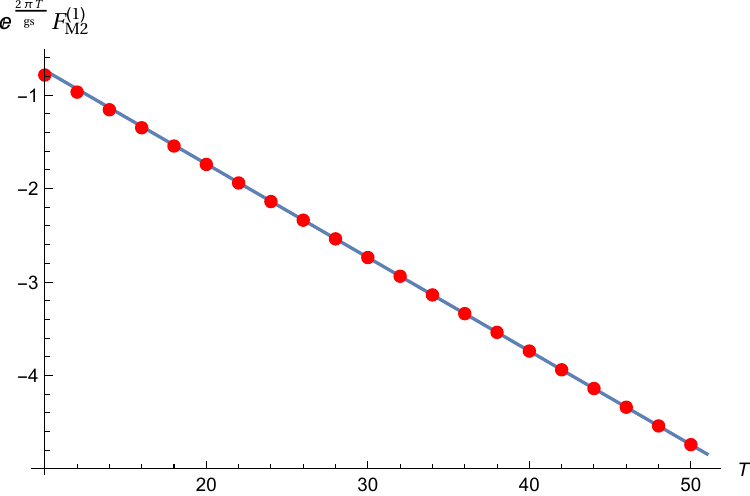}}
\hspace{15mm}
&
\resizebox{50mm}{!}{\includegraphics{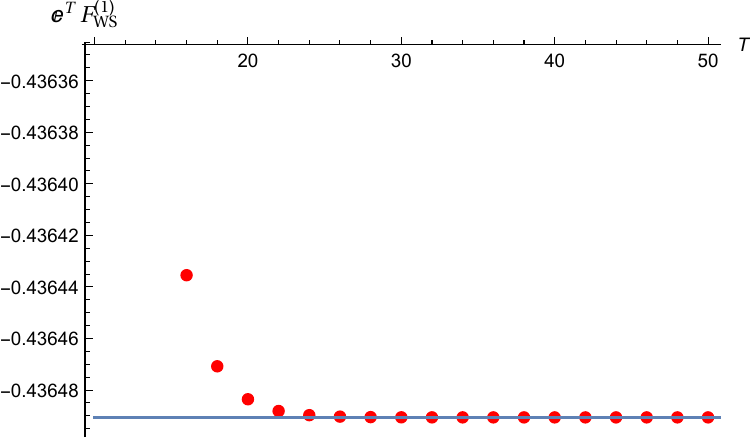}}\\
(a) Membrane 1-instanton 
\hspace{15mm}
&(b) Worldsheet 1-instanton\\
\vspace{1mm}\\
\resizebox{50mm}{!}{\includegraphics{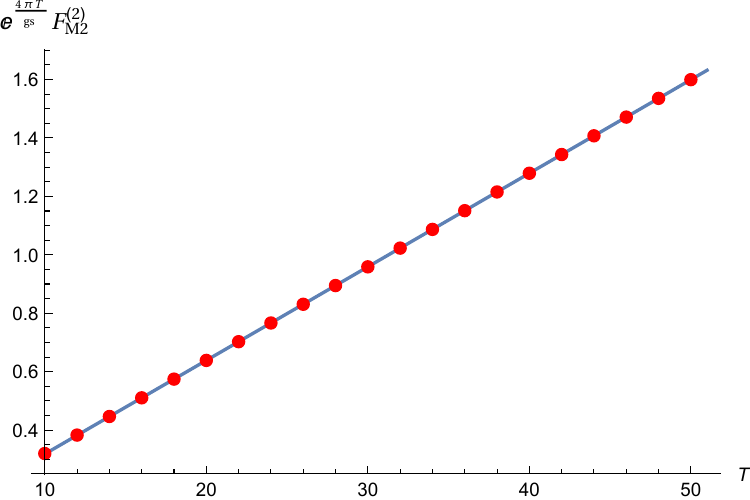}}
\hspace{15mm}
&
\resizebox{50mm}{!}{\includegraphics{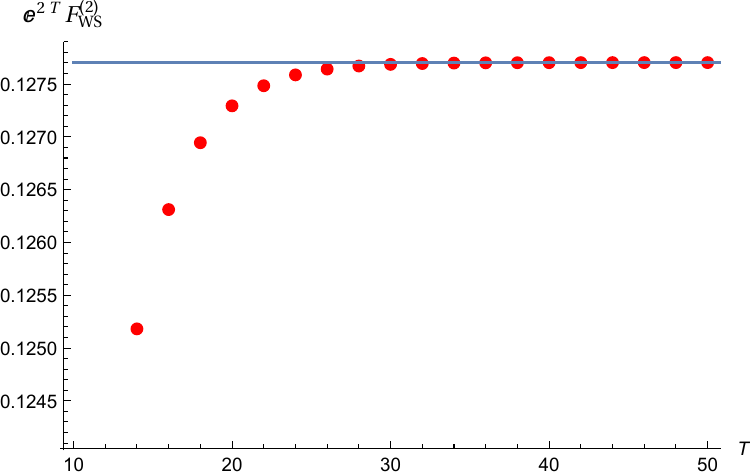}}\\
(c) Membrane 2-instanton
\hspace{15mm}
&(d) Worldsheet 2-instanton
\end{tabular}
\end{center}
\caption{We show the plots of $F_\text{coni}^\text{resum}(g_s,T+\pi\ri)$ as a function of $T$
for $g_s=8$. In the figures (a) to (d), the solid lines represent the expected worldsheet instantons
or membrane instantons
in \eqref{FM2FWSconi} normalized by the exponential factor
$\re^{-\ell T}$ or $\re^{-\frac{2\pi\ell T}{g_s}}$, while the
red dots represent the numerical values of the integral \eqref{eq:Fconi-resumT}
at each instanton order.
More precisely, the red dots in each figure represent: 
(a) $F_\text{coni}^\text{resum}(g_s,T+\pi\ri)\re^{\frac{2\pi T}{g_s}}$, 
(b) $\big[F_\text{coni}^\text{resum}(g_s,T+\pi\ri)-F_\text{M2}^{(1)}(g_s,T)\big]\re^{T}$,
(c) $\big[F_\text{coni}^\text{resum}(g_s,T+\pi\ri)-F_\text{M2}^{(1)}(g_s,T)-F_\text{WS}^{(1)}(g_s,T)\big]\re^{\frac{4\pi T}{g_s}}$,
(d) $\big[F_\text{coni}^\text{resum}(g_s,T+\pi\ri)-F_\text{M2}^{(1)}(g_s,T)-F_\text{WS}^{(1)}(g_s,T)-F_\text{M2}^{(2)}(g_s,T)\big]\re^{2T}$.
}
\label{fig:Fconi-resum}
\end{figure}
%%%%%%%%%%%%%%%%%%%%%%%%%%%%%%%%%%%%%%%%%%%%%%%%%%%%
%%%%%%%%%%%%%%%%%%%%%%%%%%%%%%%%%%%%%%%%%%%%%%%%%%%%
%%%%%%%%%%%%%%%%%%%%%%%%%%%%%%%%%%%%%%%%%%%%%%%%%%%%
%%%%%%%%%%%%%%%%%%%%%%%%%%%%%%%%%%%%%%%%%%%%%%%%%%%%
\begin{figure}[tb]
\begin{center}
\begin{tabular}{cc}
%\hspace{-3mm}
\resizebox{50mm}{!}{\includegraphics{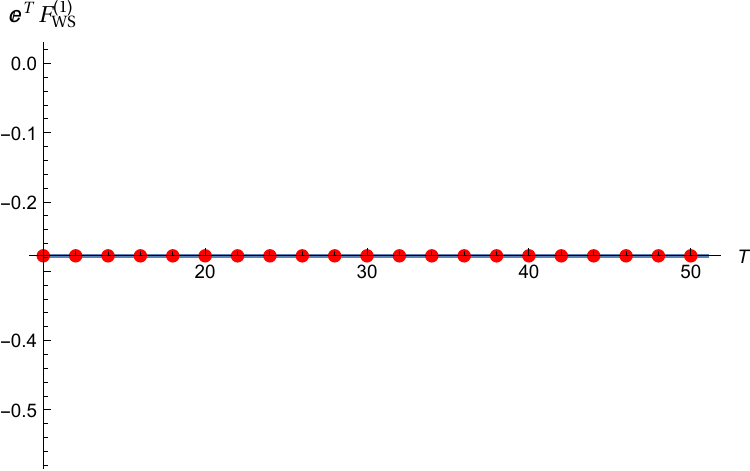}}
\hspace{15mm}
&
\resizebox{50mm}{!}{\includegraphics{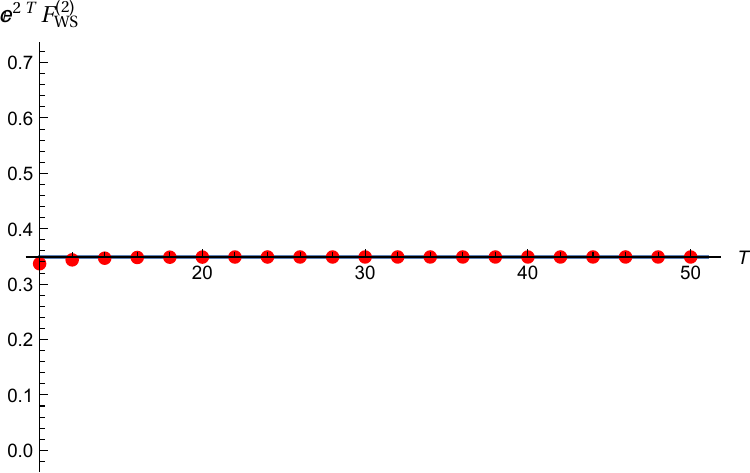}}\\
(a) Worldsheet 1-instanton 
\hspace{15mm}
&(b) Worldsheet 2-instanton\\
\vspace{1mm}\\
\resizebox{50mm}{!}{\includegraphics{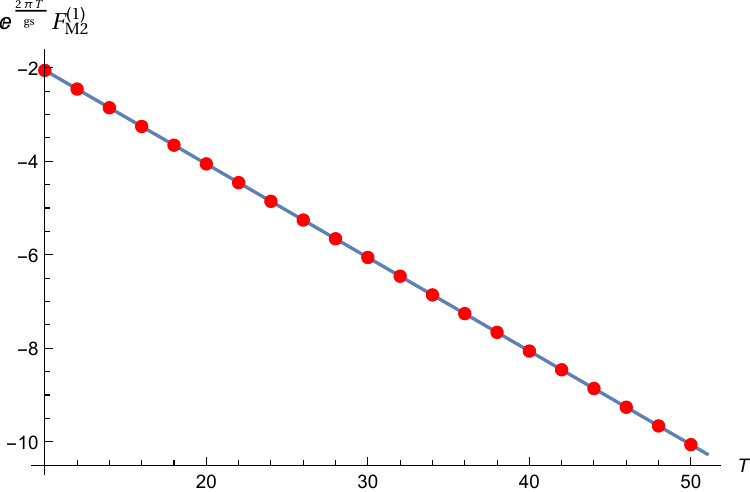}}
\hspace{15mm}
&
\resizebox{50mm}{!}{\includegraphics{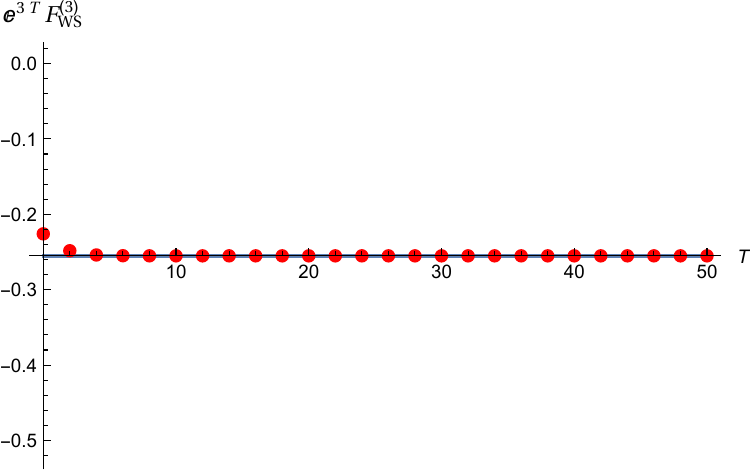}}\\
(c) Membrane 1-instanton
\hspace{15mm}
&(d) Worldsheet 3-instanton
\end{tabular}
\end{center}
\caption{We show the plots of $F_\text{coni}^\text{resum}(g_s,T+\pi\ri)$ as a function of $T$
for $g_s=2.5$. In the figures (a) to (d), the solid lines represent the expected worldsheet instantons
or membrane instantons
in \eqref{FM2FWSconi} normalized by the exponential factor
$\re^{-\ell T}$ or $\re^{-\frac{2\pi\ell T}{g_s}}$, while the
red dots represent the numerical values of the integral \eqref{eq:Fconi-resumT}
at each instanton order.
More precisely, the red dots in each figure represent: 
(a) $F_\text{coni}^\text{resum}(g_s,T+\pi\ri)\re^{T}$, 
(b) $\big[F_\text{coni}^\text{resum}(g_s,T+\pi\ri)-F_\text{WS}^{(1)}(g_s,T)\big]\re^{2T}$,
(c) $\big[F_\text{coni}^\text{resum}(g_s,T+\pi\ri)-F_\text{WS}^{(1)}(g_s,T)-F_\text{WS}^{(2)}(g_s,T)\big]\re^{\frac{2\pi T}{g_s}}$,
(d) $\big[F_\text{coni}^\text{resum}(g_s,T+\pi\ri)-F_\text{WS}^{(1)}(g_s,T)-F_\text{WS}^{(2)}(g_s,T)-F_\text{M2}^{(1)}(g_s,T)\big]\re^{3T}$.
}
  \label{fig:Fconi-resum2}
\end{figure}
%%%%%%%%%%%%%%%%%%%%%%%%%%%%%%%%%%%%%%%%%%%%%%%%%%%%
%%%%%%%%%%%%%%%%%%%%%%%%%%%%%%%%%%%%%%%%%%%%%%%%%%%%

As examples, in figure \ref{fig:Fconi-resum} and figure \ref{fig:Fconi-resum2} we show the plot of 
$F_\text{coni}^\text{resum}(g_s,T+\pi\ri)$ at $g_s=8$ and $g_s=2.5$, respectively.
In these figures, we denote the worldsheet $\ell$-instanton contribution
and the membrane $\ell$-instanton contribution in \eqref{FM2FWSconi}
as $F_\text{WS}^{(\ell)}(g_s,T)$ and
$F_\text{M2}^{(\ell)}(g_s,T)$, respectively:
\be
\ba
F_\text{WS}^{(\ell)}(g_s,T) &=\frac{(-1)^\ell}{\ell}\left(2\sin\frac{\ell g_s}{2}\right)^{-2}
\re^{-\ell T},\\
F_\text{M2}^{(\ell)}(g_s,T) &=
-\frac{1}{4\pi \ell^2}
\csc \( \frac{2\pi^2 \ell}{g_s} \) \left[
\frac{2\pi \ell}{g_s}T+\frac{2\pi^2 \ell}{g_s} \cot \( \frac{2\pi^2 \ell}{g_s} \)+1 \right]
\re^{-\frac{2\pi\ell T}{g_s}}.
\label{ell-inst}
\ea
\ee

For the $g_s=8$ case in figure \ref{fig:Fconi-resum}, the exponential factors of each instantons
have the following ordering of magnitudes
\begin{align}
 \re^{-\frac{2\pi T}{g_s}} > \re^{-T} > \re^{-\frac{4\pi T}{g_s}} > \re^{-2T}>\cdots\qquad (T\to\infty).
 \label{order-inst}
\end{align}
In particular, the first instanton correction in the large $T$
expansion is given by the membrane 1-instanton.
As we can see in figure \ref{fig:Fconi-resum}(a),
$F_\text{coni}^\text{resum}(g_s,T+\pi\ri)$
normalized by the exponential factor $\re^{-\frac{2\pi T}{g_s}}$
behaves linearly in $T$, which presicely agrees with the
coefficient of membrane 1-instanton $F_\text{M2}^{(1)}(g_s,T)$
in \eqref{ell-inst}.
Then, by subtracting the membrane 1-instanton contribution from
$F_\text{coni}^\text{resum}(g_s,T+\pi\ri)$, we find the
worldsheet 1-instanton contribution as expected from
\eqref{order-inst} (see figure \ref{fig:Fconi-resum}(b)).
By repeating this procedure, we find that
the membrane instantons
and the worldsheet instantons
indeed appear in $F_\text{coni}^\text{resum}(g_s,T+\pi\ri)$ in the expected ordering
with the correct coefficients.

On the other hand,
for the $g_s=2.5$ case in figure \ref{fig:Fconi-resum2}, the exponential factors of each instantons
have the following ordering of magnitudes
\begin{align}
  \re^{-T} > \re^{-2T}
 > \re^{-\frac{2\pi T}{g_s}} > \re^{-3T}>\cdots\qquad (T\to\infty).
 \label{order-inst2}
\end{align}
Again, as we can see in figure \ref{fig:Fconi-resum2},
 the worldsheet instantons
 and the membrane instantons
appear in $F_\text{coni}^\text{resum}(g_s,T+\pi\ri)$ in the expected ordering \eqref{order-inst2}.
We have also checked numerically for some other values of $g_s$
that $F_\text{coni}^\text{resum}(g_s,T+\pi\ri)$
correctly includes the worldsheet instantons and the membrane instantons.

Furthermore, 
if $g_s/\pi$ is an integer, 
one can write down the right hand side in \eqref{eq:F-coni-resum2} in a closed form.
For such a value, both the Gopakumar-Vafa formula \eqref{eq:F-coni} and the non-perturbative
correction \eqref{eq:F-coni-np} are divergent, but the sum is totally finite,
as mentioned in the introduction.
For example, for $g_s=2\pi$, one obtains
\be
\ba
&F_\text{coni}^\text{GV}(g_s,T+\pi\ri)+F_\text{coni}^\text{np}(g_s,T)|_{g_s=2\pi}
= \sum_{\ell=1}^\infty \frac{(-1)^\ell}{8\pi^2} \( \frac{T^2}{\ell}+\frac{2T}{\ell^2}
+\frac{2}{\ell^3}+\frac{\pi^2}{\ell} \)\re^{-\ell T} \\
&\qquad=\frac{1}{4\pi^2} \Li_3 (-\re^{-T})+\frac{T}{4\pi^2} \Li_2(-\re^{-T})-\( \frac{T^2}{8\pi^2}+\frac{1}{8} \)
\log (1+\re^{-T}).
\ea
\label{eq:F-coni-2pi}
\ee
We numerically confirmed that this expression reproduces the resummed result \eqref{eq:Fconi-resumT}
for various values of $T$.
Similarly, for $g_s=\pi$, we find
\be
\ba
F_\text{coni}^\text{GV}(g_s,T+\pi\ri)+F_\text{coni}^\text{np}(g_s,T)|_{g_s=\pi}&=\frac{1}{8\pi^2} \Li_3 (\re^{-2T})
+\frac{T}{4\pi^2} \Li_2(\re^{-2T})\\
&-\( \frac{T^2}{4\pi^2}+\frac{1}{8} \)\log (1-\re^{-2T})
-\frac{1}{4}\arctanh \re^{-T}.
\ea
\label{eq:F-coni-pi}
\ee
We observe that for $g_s=n \pi$ ($n \in \mathbb{Z}$), 
the right hand side in \eqref{eq:F-coni-resum2} can also be
written as a closed form.%
\footnote{The exact expressions \eqref{eq:F-coni-2pi} and \eqref{eq:F-coni-pi} are very
similar to the ones for the grand potential in ABJM theory at $k=2$ and $k=4$, respectively,
in \cite{CGM, GHM2}.
In fact, these expressions can be written as
\begin{align*}
F_\text{coni}^\text{GV}(g_s,T+\pi\ri)+F_\text{coni}^\text{np}(g_s,T)|_{g_s=2\pi}
&=\frac{1}{4\pi^2}\( F_0^\text{inst}-T \pd_T F_0^\text{inst}+\frac{1}{2} T^2 \pd_T^2 F_0^\text{inst} \)
+\frac{3}{2}F_1^\text{inst}, \\
F_\text{coni}^\text{GV}(g_s,T+\pi\ri)+F_\text{coni}^\text{np}(g_s,T)|_{g_s=\pi}
&=\frac{1}{8\pi^2}\( F_0^\text{inst}-T \pd_T F_0^\text{inst}+\frac{1}{2} T^2 \pd_T^2 F_0^\text{inst} \)
+\frac{3}{2}F_1^\text{inst}-\frac{1}{4}\arctanh \re^{-T},
\end{align*}
where $F_g^\text{inst}=F_g^\text{inst}(T+\pi \ri)$ for $g_s=2\pi$ and 
$F_g^\text{inst}=F_g^\text{inst}(2T)$ for $g_s=\pi$ with $F_0^\text{inst}(t)=\Li_3(\re^{-t})$ and 
$F_1^\text{inst}(t)=\frac{1}{12} \Li_1(\re^{-t})$.
}
The result is in perfect agreement with the resummed formula \eqref{eq:Fconi-resumT}
even for small $T$.
We conclude that the resummation formula \eqref{eq:Fconi-resumT} indeed contains
both the worldsheet instanton correction $F_\text{coni}^\text{GV}(g_s,T+\pi \ri)$
and the membrane instanton correction $F_\text{coni}^\text{np}(g_s,T)$.

\subsection{Comparison to the standard Borel resummation}
Here we check that our resummation \eqref{eq:F-coni-resum} or \eqref{eq:Fconi-resumT} 
by the Bernoulli moment method
is equivalent to the standard Borel resummation.
Let us consider the Borel resummation of the genus expansion \eqref{eq:FGVexp}.
In particular, we focus on the genus $g \geq 2$ contribution:
\be
\widetilde{F}_\text{coni}^\text{GV}(g_s,t)=\sum_{g=2}^\infty g_s^{2g-2}\til{F}_g(t),\qquad
\til{F}_{g \geq 2}(t)=
\frac{(-1)^{g-1}B_{2g}}{2g(2g-2)!}\Li_{3-2g}(\re^{-t}).
\label{eq:Fg-all-genus}
\ee
In the following, we use a bit modified Borel transform rather than \eqref{eq:Borel-transform},
following \cite{PS}:
\be
\cB [\widetilde{F}_\text{coni}^\text{GV}](\zeta)
=\sum_{g=2}^\infty \frac{\zeta^{2g-2}}{(2g-3)!}\widetilde{F}_g(t),
\label{eq:BF-coni}
\ee
Then the Borel resummation of $\widetilde{F}_\text{coni}^\text{GV}$ along the positive real axis 
is given by
\be\ba
\cS_0\widetilde{F}_\text{coni}^\text{GV}(g_s,t)
=\int_0^\infty \frac{\rd \zeta}{\zeta}\, \re^{-\zeta} \cB [\widetilde{F}_\text{coni}^\text{GV}](g_s \zeta)
=\int_0^\infty \frac{\rd \zeta}{\zeta} \, \re^{-\zeta/g_s} \cB [\widetilde{F}_\text{coni}^\text{GV}](\zeta).
\ea\ee
The first important step is to understand the singularity structure of the Borel transform 
$\cB [\widetilde{F}_\text{coni}^\text{GV}](\zeta)$.
It turns out that the singularity structure of 
$\cB [\widetilde{F}_\text{coni}^\text{GV}](\zeta)$
heavily depends on $t$.
\begin{figure}[tb]
\begin{center}
\begin{tabular}{ccc}
%\hspace{-3mm}
\resizebox{46mm}{!}{\includegraphics{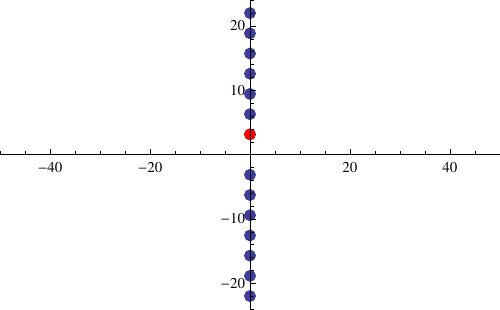}}
%\hspace{0mm}
&
\resizebox{46mm}{!}{\includegraphics{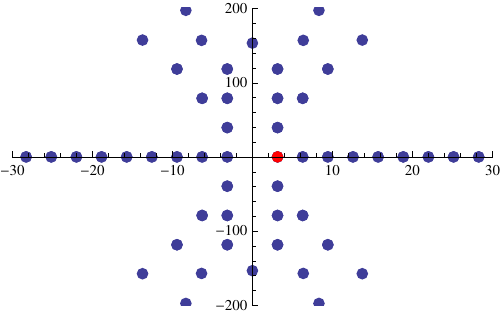}}
%\vspace{-5mm}
&
\resizebox{46mm}{!}{\includegraphics{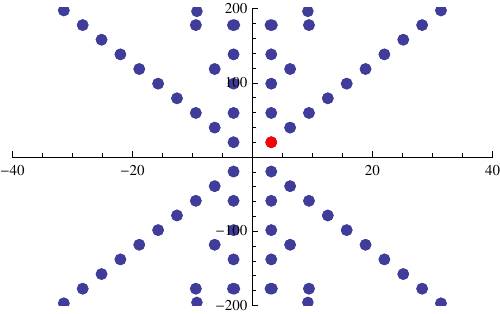}}
%\vspace{-5mm}
\end{tabular}
\end{center}
  \caption{We show the singularities of $\cB [\widetilde{F}_\text{coni}^\text{GV}](\zeta)$ 
in the Borel plane for
 $t=\frac{\ri}{2}$ (Left), $t=\frac{1}{2}$ (Middle) and $t=\frac{1}{2}+\pi \ri$ (Right). 
For $t \in \mathbb{R}$, there exist
  an infinite number of singularities
 on the positive real axis, and thus the genus expansion is non-Borel summable.  
 For $t \in \mathbb{R}+\pi \ri$, on the other hand, no singularities lie on the positive real axis.
In all the cases, one of the closest singularities to the origin is located at $\zeta=2\pi t$, which we show by the red points.}
  \label{fig:Borel-sing-coni}
\end{figure}
In figure~\ref{fig:Borel-sing-coni}, we show the singularities of $\cB [\widetilde{F}_\text{CS}](\zeta)$
for $t=\frac{\ri}{2}, \frac{1}{2}, \frac{1}{2}+\pi \ri$.
As in the left figure, if $t \in \ri \mathbb{R}$, which corresponds to the physical Chern-Simons case,
no singularities lie on the positive real axis. 
The middle figure shows that for $t \in \mathbb{R}$, there are an infinite number of singularities on 
the positive real axis.
Therefore in this case the genus expansion is \textit{not Borel summable}.
This case was analyzed in \cite{PS} in detail.
We will also revisit this case in the next subsection.
On the other hand, in the case of \eqref{eq:t-B}, there are no singularities on the positive real axis
(see the right figure), and thus the genus expansion is \textit{Borel summable.}
This is a crucial difference from the setup in \cite{PS}.
As shown by the red points in figure~\ref{fig:Borel-sing-coni}, 
one of the closest singularities to the origin is located at $\zeta=2\pi t$.

Now let us compare our resummation \eqref{eq:F-coni-resum} with
the standard Borel resummation:
\be
\cS_0 F_\text{coni}^\text{GV}(g_s,t)
=\frac{\text{Li}_3(\re^{-t})}{g_s^2}+
\frac{1}{12}\text{Li}_1(\re^{-t})+\cS_0\widetilde{F}_\text{coni}^\text{GV}(g_s,t).
\label{eq:F-coni-Borel}
\ee
As an example, let us see the case of $t=\frac{1}{2}+\pi \ri$ in detail.
The comparison for other values of $t$ is straightforward.
In the practical computation, we replace the Borel transform 
$\cB [\widetilde{F}_\text{coni}^\text{GV}](\zeta)$ by its Pad\'e approximant.
Let us consider the Pad\'e approximation of a given function
\be
f(x)=\sum_{n=0}^\infty f_n x^n.
\ee
We denote its Pad\'e approximant with numerator order $M$ and denominator order $N$ by
\be
f^{[M/N]}(x)=\frac{a_0+a_1 x+\cdots a_M x^M}{1+b_1 x+\cdots b_N x^N}.
\ee 
The $M+N+1$ coefficients $\{ a_0,\dots,a_M;b_1,\dots,b_N \}$ are uniquely fixed by requiring 
\be
f(x)-f^{[M/N]}(x)=\cO(x^{M+N+1}).
\ee
The special case $f^{[M/M]}(x)$ is called the diagonal Pad\'e approximant with order $M$.
Using the Pad\'e approximant, we define the Borel-Pad\'e resummation by
\be\ba
\cS_0^{[M/N]} \widetilde{F}_\text{coni}^{\text{GV}}(g_s,t)
=\int_0^\infty \frac{\rd \zeta}{\zeta} \, \re^{-\zeta/g_s} \cB [\widetilde{F}_\text{coni}^{\text{GV}}]^{[M/N]}(\zeta).
\label{eq:F-coni-Borel-Pade}
\ea\ee
It is natural to expect that the Borel-Pad\'e resummation converges to the Borel resummation
in the limit $M \to \infty$ and $N \to \infty$.
Since we know the all-genus result \eqref{eq:Fg-all-genus}, we can compute the Pad\'e approximant
up to any desired order.
In table~\ref{tab:resum}, we show the numerical values of  $\cS_0^{[200/200]} F_\text{coni}^\text{GV}(g_s,t)$
and the difference $|\cS_0^{[200/200]} F_\text{coni}^\text{GV}-F_\text{coni}^\text{resum}|$ for various 
finite $g_s$ with $t=\frac{1}{2}+\pi \ri$.
These two resummations show a very good agreement.
Of course, one can push the same computation for other values of $t$.

\begin{table}[tb]
\caption{The comparison of the Borel-Pad\'e resummation \eqref{eq:F-coni-Borel} with \eqref{eq:F-coni-Borel-Pade} and our 
resummation \eqref{eq:F-coni-resum}.
The K\"ahler modulus is set to be $t=\frac{1}{2}+\pi \ri$. 
In the evaluation of the Borel resummation, we use the diagonal Pad\'e approximant 
$\cB [\widetilde{F}_\text{coni}^\text{GV}]^{[200/200]}(\zeta)$.
Note that the original Gopakumar-Vafa formula \eqref{eq:F-coni} has singularities for rational $g_s/\pi$,
but  the resummations \eqref{eq:F-coni-Borel} or \eqref{eq:Fconi-resumT} do not have them any more.}
\label{tab:resum}
\begin{center}
  \begin{tabular}{ccc}\hline
$g_s$ & $\cS_0^{[200/200]} F_\text{coni}^\text{GV}(g_s, t=\frac{1}{2}+\pi \ri)$ & $|\cS_0^{[200/200]} F_\text{coni}^\text{GV}-F_\text{coni}^\text{resum}|$ \\ \hline  
$\pi$ & $-0.1055363639228783289891108649454531297931$ & $< 10^{-40}$ \\
$2\pi$    & $-0.08188090730148760004781509906298147491485$ & $< 10^{-40}$ \\
$3\pi$ & $-0.09808972547164339586210530022543766417197$ & $< 10^{-40}$ \\
$4\pi$    & $-0.1217659634537775320983860936659306511076$ & $< 10^{-40}$ \\
$5\pi$ & $-0.1481839809776953950785808068127348899025$ & $< 10^{-40}$ \\
$6\pi$    & $-0.1759365690980661891778399697297576048860$ & $5.1 \times 10^{-38}$ \\
\hline
\end{tabular}
\end{center}
\end{table}

\begin{figure}[tb]
\begin{center}
\begin{tabular}{cc}
%\hspace{-3mm}
\resizebox{70mm}{!}{\includegraphics{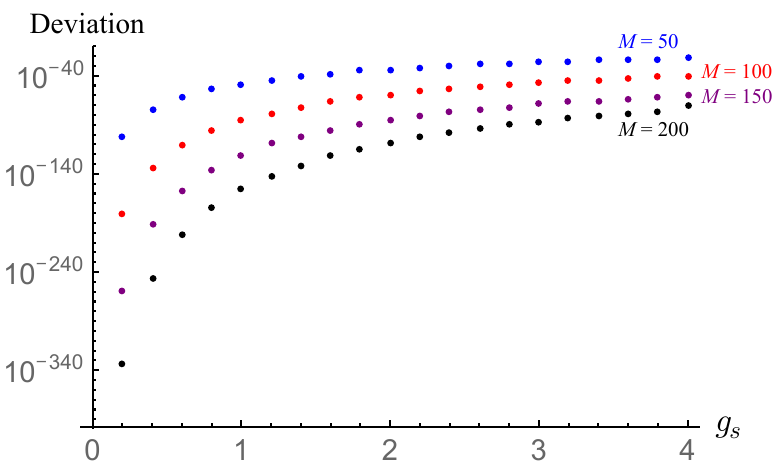}}
%\hspace{10mm}
&
\resizebox{70mm}{!}{\includegraphics{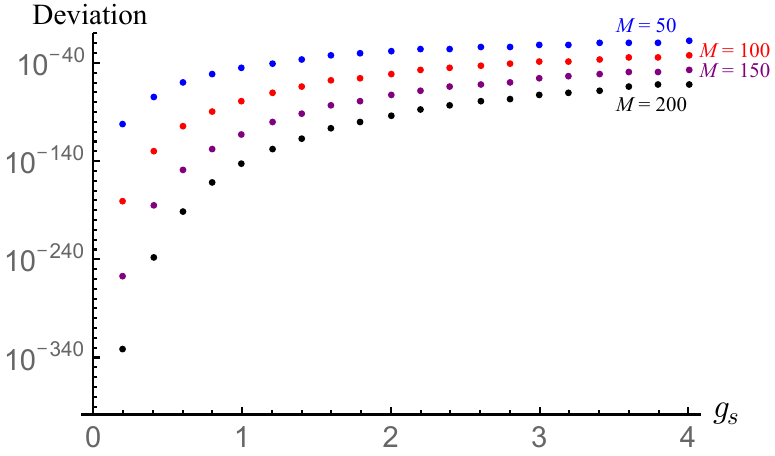}}
\vspace{2mm}
\\
{\footnotesize $t=1+\pi \ri$} &
{\footnotesize $t=3+\pi \ri$}
\end{tabular}
\end{center}
\vspace{-3mm}
  \caption{The deviation 
$|(\cS_0^{[M/M]} F_\text{coni}^\text{GV}-F_\text{coni}^\text{resum})
/F_\text{coni}^\text{resum}|$ is plotted
against the string coupling $g_s$.
We shows the graphs for $t=1+\pi \ri, 3+\pi \ri$ and for $M=50,100,150,200$.
It is clear to see that the deviation gets smaller as the order $M$ of the Pad\'e approximant
gets larger.
}
  \label{fig:deviation-gs}
\end{figure}

\begin{figure}[tb]
\vspace{7mm}
\begin{center}
\begin{tabular}{cc}
\resizebox{70mm}{!}{\includegraphics{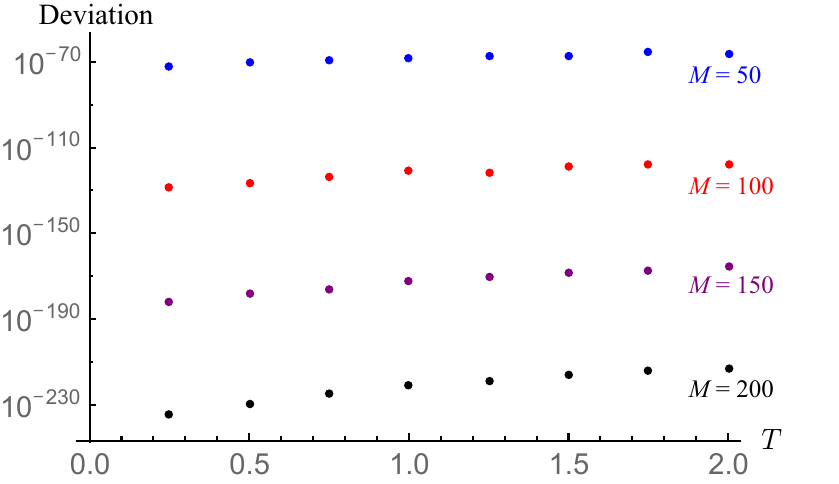}}
%\hspace{10mm}
&
\resizebox{70mm}{!}{\includegraphics{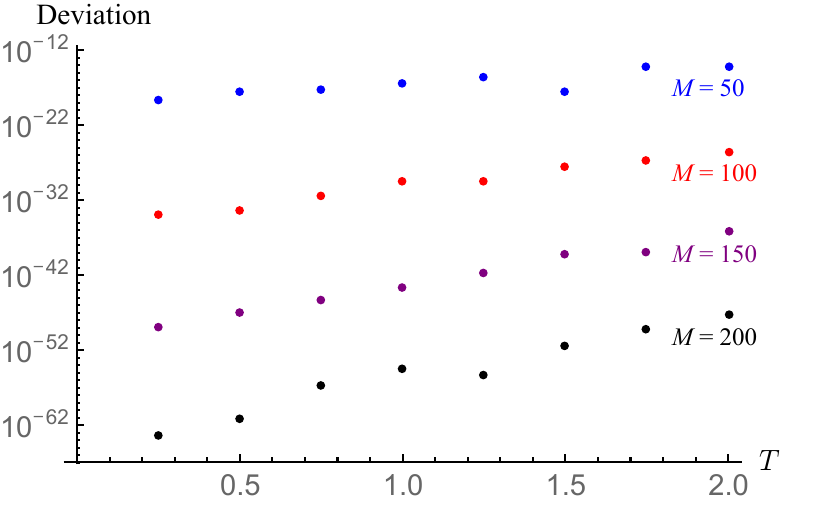}}
\vspace{2mm}
\\
{\footnotesize $g_s=1/2$} &
{\footnotesize $g_s=7$}
\end{tabular}
\end{center}
\vspace{-3mm}
  \caption{The deviation 
$|(\cS_0^{[M/M]} F_\text{coni}^\text{GV}-F_\text{coni}^\text{resum})
/F_\text{coni}^\text{resum}|$ is plotted
against the mudulus $t=T+\pi \ri$.
We shows the graphs for $g_s=1/2, 7$ and for $M=50,100,150,200$.
}
  \label{fig:deviation-t}
\end{figure}

In figure~\ref{fig:deviation-gs}, we plot the deviation 
$|(\cS_0^{[M/M]} F_\text{coni}^\text{GV}-F_\text{coni}^\text{resum})/F_\text{coni}^\text{resum}|$
for fixed $t=1+\pi \ri, 3+\pi \ri$ as a function of the string coupling $g_s$.
We used the several diagonal Pad\'e approximants with order $M=50,100,150,200$.
These graphs show that the Borel-Pad\'e resummation indeed gets closer to the our resummation
as the order of the Pad\'e approximant gets larger.
In figure~\ref{fig:deviation-t}, we similarly show the deviation 
for fixed $g_s=1/2,7$ as a function of the K\"ahler modulus $t=T+\pi \ri$.
These numerical tests strongly suggest that the standard Borel resummation and our resummation
\eqref{eq:F-coni-resum}
 are exactly equivalent:
\be
\cS_0 F_\text{coni}^\text{GV}(g_s,t)=F_\text{coni}^\text{resum}(g_s,t).
\ee

\subsection{Non-Borel summable case revisited}\label{sec:non-Borel}
\begin{figure}[tb]
\begin{center}
\resizebox{75mm}{!}{\includegraphics{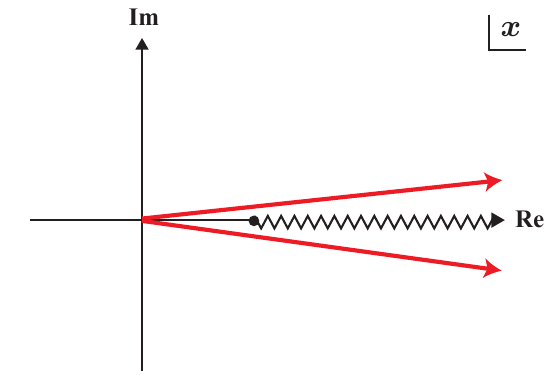}}
\end{center}
\vspace{-0.3cm}
\caption{When $t$ is real, the integrand in \eqref{eq:F-coni-resum} has a branch cut 
along $x\geq |t/g_s|$. One has to deform the integration contour in order to avoid this branch cut.
As shown by the red arrows, there are two such deformations, and these two lead to different lateral resummations \eqref{eq:F-coni-resum-lateral}.}
  \label{fig:Lateral-resum}
\end{figure}

As mentioned before, if
$t$ is real,
the resummation formula \eqref{eq:F-coni-resum} is not well-defined.
One has to deform the integration contour to avoid the singularity in the integrand.
This situation just corresponds to the fact that the genus expansion \eqref{eq:FGVexp}
is non-Borel summable for $t \in \mathbb{R}$ (see the middle in figure~\ref{fig:Borel-sing-coni}).
The Borel analysis in this case was performed in \cite{PS} in great detail.
Here we revisit this case in order to use our resummation formula \eqref{eq:F-coni-resum}.
As shown in figure~\ref{fig:Lateral-resum}, one has to avoid the branch cut $(|t/g_s|, \infty)$ 
in the integrand in \eqref{eq:F-coni-resum}.
These contour deformations lead to the following two \textit{lateral resummations}:
\be
F_\pm^\text{resum}(g_s,t)=\frac{\Li_3(\re^{-t})}{g_s^2}
-\int_0^{\infty \re^{\pm \ri 0}} \rd x \frac{x}{\re^{2\pi x}-1} \log (1+\re^{-2t}-2\re^{-t} \cosh g_s x ).
\label{eq:F-coni-resum-lateral}
\ee
These resummed free energies are no longer real-valued,
and have a discontinuity
\be
\Disc F^\text{resum}(g_s,t)=F_+^\text{resum}(g_s,t)-F_-^\text{resum}(g_s,t) \ne 0.
\ee
%%%%%%%%%%%%%
%%%%%%%%%%%%%
We observe that these lateral resummations are still written as the form
\be
F_\pm^\text{resum}(g_s,t)=F_\text{coni}^\text{GV}(g_s,t)+F_\text{coni}^\text{np}(g_s,t \pm \pi \ri),
\label{eq:F-coni-resum-lateral2}
\ee
where the non-perturbative part $F_\text{coni}^\text{np}(g_s,t)$ is given by \eqref{eq:F-coni-np}.
For instance, if we set $g_s=t=1$, the integral \eqref{eq:F-coni-resum-lateral} leads to
\be
F_\pm^\text{resum}(1,1)\approx 0.43025505451123409079 \pm 0.00108327009532842310 \ri .
\label{eq:F-pm-special}
\ee 
Each term on the right hand side in \eqref{eq:F-coni-resum-lateral2} can be evaluated as%
\footnote{For $g_s=1$, both the Gopakumar-Vafa part and the non-perturbative part are finite
because $g_s/\pi$ is not rational.
}
\be
\ba
F_\text{coni}^\text{GV}(1,1) &\approx 0.43599560928729258893, \\
F_\text{coni}^\text{np}(1,1 \pm \pi \ri) &\approx 
-0.00574055477605849814 \pm 0.00108327009532842310 \ri.
\ea
\ee
The sum of these two contributions precisely agrees with \eqref{eq:F-pm-special}.

Now let us compute the discontinuity of \eqref{eq:F-coni-resum-lateral} exactly.
It is obvious that the first term in \eqref{eq:F-coni-resum-lateral2} is real for $t \in \mathbb{R}$
and $g_s \in \mathbb{R}$.
Thus the imaginary part comes from the non-perturbative part.
It is not hard to compute the imaginary part of $F_\text{coni}^\text{np}(g_s,t \pm \pi \ri)$.
One finds
\be
\im [ F_\text{coni}^\text{np}(g_s,t \pm \pi \ri)]
=\pm \left[ \frac{1}{4\pi} \Li_2(\re^{-\frac{2\pi t}{g_s}} )-\frac{t}{2g_s} \log(1-\re^{-\frac{2\pi t}{g_s}} ) 
\right].
\ee
Therefore the discontinuity of \eqref{eq:F-coni-resum-lateral} is finally given by
\be
\Disc F^\text{resum}(g_s,t)=
\frac{\ri}{2\pi} \left[ \Li_2(\re^{-\frac{2\pi t}{g_s}} )
-\frac{2\pi t}{g_s} \log(1-\re^{-\frac{2\pi t}{g_s}} ) \right]  .
\label{eq:DiscF}
\ee
Of course, the same result should be derived directly from the integral representations
\eqref{eq:F-coni-resum-lateral}.
We confirmed it numerically.
One notices that this discontinuity is exactly the same as that for the $c=1$ strings 
at self-dual radius (see (4.6) in \cite{PS}).
In fact, it was noted in \cite{PS} that the Borel singularities on the positive real axis 
in the genus expansion \eqref{eq:FGVexp} come from 
the $c=1$ string free energy.
Our resummation formula \eqref{eq:F-coni-resum-lateral} is perfectly consistent with
this fact.
%\footnote{The discontinuity of the resolved conifold free energy in \cite{PS} looks a little bit different from this.
%It was observed in \cite{Hatsuda} that their result is almost the same as 
%$F_\text{coni}^\text{np}(g_s,t + \pi \ri)$ itself, not its imaginary part.
%}
In other words, the difference $F_\text{coni}-F_{c=1}$ has no singularities on the positive real axis
in the Borel plane, and is Borel summable. 

Let us give a remark on the comparison to the discontinuity formula in \cite{PS}.
In \cite{PS}, the discontinuity for the resolved conifold free energy was also computed (see (4.23) in \cite{PS}).
Their result looks quite different from ours \eqref{eq:DiscF}.
Why? The reason is the following.
In \cite{PS}, it was confirmed that their discontinuity reproduces the all-order genus expansion
after using the Cauchy formula.
The important point is that to reproduce the original perturbative asymptotic expansion from the Cauchy formula,
we need the information on \textit{all the singularities on the Borel plane}.
As shown in the middle in figure~\ref{fig:Borel-sing-coni}, the Borel transform \eqref{eq:BF-coni} has an infinite number of complex singularities in addition to the ones on the positive real axis.
The fact that the discontinuity in \cite{PS} reproduces the correct genus expansion
strongly suggests that their discontinuity formula includes the information on the discontinuities 
along all the singular directions.
On the other hand, we here considered the two lateral Borel resummations deformed from the positive real axis.
As shown in figure~\ref{fig:Lateral-resum}, it is obvious to see that the discontinuity of these two lateral Borel resummations
is caused by the \textit{singularities only on the positive real axis}.
These singularities come from the $c=1$ string free energy, as noted in \cite{PS}.
Our discontinuity \eqref{eq:DiscF} just corresponds to the $m=0$ term in (4.23) in \cite{PS}.
Of course, the $m=0$ term alone reproduces the genus expansion of the $c=1$ string free energy, 
not of the full resolved conifold free energy.
We conclude that the contributions for $m \ne 0$ in (4.23) in \cite{PS} should come from the other singularities on the Borel plane.

The discontinuity \eqref{eq:DiscF} causes an ambiguity of the lateral resummations.
This ambiguity must be canceled by additional non-perturbative corrections 
to \eqref{eq:F-coni-resum-lateral2}
because the complete free energy should be an \textit{unambiguous} quantity.
Thus it is natural to consider a trans-series expansion.
However, we emphasize that the resummations \eqref{eq:F-coni-resum-lateral2} already
have the non-perturbative correction $F_\text{coni}^\text{np}$ in the M-theory expansion: $t \to \infty$ with finite $g_s$, which contributes
to the real part of the resummations.
To explore the trans-series expansion is beyond the scope of this work.
It would be interesting to ask what the non-perturbative sector in the trans-series implies in the M-theory expansion.

\section{Comments on the Borel resummation of ABJM free energy}\label{sec:ABJM-resum}
In this section, we revisit the Borel resummation of the free energy of 
the $U(N)_k\times U(N)_{-k}$ ABJM theory
on a three-sphere,
which is related to the topological string on the ``diagonal'' 
local $\mathbb{P}^1\times \mathbb{P}^1$, where the K\"{a}hler parameters of two  $\mathbb{P}^1$'s are equal.
In the ABJM theory, the non-perturbative corrections
are well understood \cite{DMP2,MP2,HMO2,CM,HMO3}, and in particular
it is known that the 
membrane instanton corrections are given by the refined topological string in
the Nekrasov-Shatashvili limit, on the same local $\mathbb{P}^1\times \mathbb{P}^1$ \cite{HMMO}.
The resummation of the genus expansion in this case has been already studied in \cite{GMZ}.
For the physical parameters of ABJM theory,
the genus expansion is Borel summable, 
but as observed in \cite{GMZ} the Borel resummation does not reproduce the correct 
membrane instantons in ABJM theory.  
In \cite{GMZ}, the missing information is attributed to
the effect of so-called complex instantons \cite{DMP2}.

Here we would like to show that the Borel resummation of
ABJM free energy contains membrane-like non-perturbative corrections which 
removes the singularity in the Gopakumar-Vafa formula of the 
worldsheet instanton part of the free energy.
These membrane-like corrections are not equal to 
the complete membrane instantons in the ABJM theory given by
the Nekrasov-Shatashvili limit, but they have a similar form with the
membrane instantons in the resolved conifold case, as we will see below.

As shown in \cite{DMP1},
the genus expansion of the free energy of ABJM theory 
can be systematically computed by solving
the holomorphic anomaly equation.
For this purpose it is useful to write
everything in term of the modular parameter $\tau$
of the spectral curve of the ABJM matrix model
\be\ba
 \tau&=\frac{\ri}{4\pi^3}\del_\la^2F_0(\la)+1=\ri \frac{K'(\ri\kappa/4)}{K(\ri\kappa/4)},\\
\la&=\frac{\kappa}{8\pi}{}_3F_2\left(\hf,\hf,\hf;1,\frac{3}{2};-\frac{\kappa^2}{16}\right),
\ea\ee
where $\la=N/k$ is the 't Hooft parameter of ABJM theory
and $F_0(\la)$ is the genus-zero part of the free energy
\begin{align}
 F=\log Z_\text{ABJM}(N,k)=\sum_{g=0}^\infty F_g(\la)\left(\frac{2\pi \ri}{k}\right)^{2g-2}.
\end{align}
The genus-one free energy is given by
\begin{align}
 F_1=-\log\eta(\tau)-\hf\log 2+2\zeta'(-1)+\frac{1}{6}\log\frac{4\pi}{k}.
\end{align}
To find the higher genus corrections,
we first introduce $f_g$ by
\begin{align}
 F_g=\xi^{2g-2}f_g,
\end{align}
with
\begin{align}
\xi=\frac{2}{b^\hf d},\quad b=\th_2^4(\tau),\quad
d=\th_4^4(\tau).
\end{align}
Then the holomorphic anomaly equation becomes
\begin{align}
 -3\frac{\rd f_g}{\rd E_2}=\rd_\xi^2 f_{g-1}+\frac{\del_\tau\xi}{3\xi}\rd_\xi f_{g-1}
+\sum_{r=1}^{g-1}\rd_\xi f_{r}\rd_\xi f_{g-r},
\label{eq:HAE}
\end{align}
where $E_2(\tau)$ is the Eisenstein series and
$\rd_\xi$ denotes the covariant derivative
which sends a modular form of weight $k$ to  weight $k+2$
\begin{align}
 \rd_\xi=\del_\tau+\frac{k\del_\tau\xi}{3\xi}.
\end{align}
Starting from $\rd_\xi f_1=-E_2/24$,
one can solve \eqref{eq:HAE} recursively.
The integration constant, known as the holomorphic ambiguity, can be fixed
by the so-called gap condition
at the conifold point and the orbifold point (see \cite{DMP1} for the detail).
In this way, we have computed  the genus $g$ free energy $F_g$
up to $g=101$.
For example, the genus-$2$ and genus-$3$ free energies are given by
\begin{align}
 F_2&=-\frac{5 E_2^3}{1296 b d^2}+\frac{E_2^2}{144
   d^2}+
\frac{E_2 \left(-b^2-b d-d^2\right)}{216 b d^2}+\frac{16 b^3+15 b^2 d-15 b
   d^2+2 d^3}{12960 b d^2},\nn
F_3&=
\frac{5 E_2^6}{46656 b^2 d^4}
-\frac{25 E_2^5}{46656 b d^4}
+\frac{E_2^4 \left(20 b^2+10 b d+3
   d^2\right)}{11664 b^2 d^4}
+\frac{E_2^3 \left(-529 b^3-519 b^2 d-111 b d^2+52
   d^3\right)}{139968 b^2 d^4}\nn
&\quad+\frac{E_2^2 \left(1172 b^4+1772 b^3 d+825 b^2
   d^2+208 b d^3+163 d^4\right)}{233280 b^2 d^4}\nn
&\quad+\frac{E_2 \left(-844 b^5-1689
   b^4 d-1205 b^3 d^2-302 b^2 d^3+102 b d^4+104 d^5\right)}{233280 b^2
   d^4}\nn
&+\frac{10718 b^6+26340 b^5 d+24699 b^4 d^2+11990 b^3 d^3+3135 b^2 d^4+1320 b
   d^5+1016 d^6}{9797760 b^2 d^4}.
\end{align}
Using the data of genus expansion, we can consider the Borel resummation
\begin{align}
\mathcal{S}_0 F=
 F_0\left(\frac{2\pi \ri}{k}\right)^{-2}+F_1+\int_0^\infty \frac{\rd\zeta}{\zeta}\re^{-\zeta}
\sum_{g=2}^\infty \frac{F_g}{(2g-3)!}\left(\frac{2\pi \ri \zeta}{k}\right)^{2g-2}.
\end{align}
As observed in \cite{GMZ},
for the physical case $\la,k\in\mathbb{R}$
there are no poles along the positive real axis on the Borel $\zeta$-plane, and
the genus expansion of ABJM theory is Borel summable.
However, the resummed free energy $\mathcal{S}_0 F$ does not agree with
the exact free energy of ABJM theory.
To see the difference, it is convenient, as in the Fermi-gas formalism, to introduce
the approximated grand potential $J_\text{Borel}(\mu)$
associated with
the Borel sum $\mathcal{S}_0 F$
\begin{align}
 \re^{\mathcal{S}_0 F}=\int_C\frac{\rd\mu}{2\pi\ri}\,\re^{J_\text{Borel}(\mu)-N\mu},
\label{Jborelint}
\end{align}
where the contour $C$ starts from $e^{-\frac{\pi\ri}{3}}\infty$ and goes to
$e^{+\frac{\pi\ri}{3}}\infty$ in the $\mu$-plane.
From the numerical analysis using the Borel-Pad\'e approximation, we find that
$J_\text{Borel}(\mu)$ has the following structure:
\begin{align}
 J_\text{Borel}(\mu)=J_\text{pert}(\mu)+
J_\text{WS}(\mu)+J_\text{np}(\mu).
\label{JBorel}
\end{align}
Here $J_\text{pert}(\mu)$ in \eqref{JBorel} is the perturbative part
\begin{align}
 J_\text{pert}(\mu)=\frac{C(k)\mu^3}{3}+B(k)\mu+A_\text{c}(k),
\end{align}
where $A_\text{c}(k)$ is given by \eqref{eq:A-int}
and $C(k)$ and $B(k)$ are given by
\begin{align}
 C(k)=\frac{2}{\pi^2k},\quad B(k)=\frac{1}{3k}+\frac{k}{24},
\end{align}
and $J_\text{WS}(\mu)$ in \eqref{JBorel} is
the worldsheet instanton corrections
\begin{align}
J_\text{WS}(\mu)=\sum_{n=1}^\infty \sum_{d=1}^\infty\sum_{g=0}^\infty 
n_g^d \(2\sin\frac{2\pi n}{k}\)^{2g-2}\frac{(-1)^{dn}\re^{-\frac{4dn\mu}{k}}}{n},
\label{JGV}
\end{align}
where $n_g^d$ is the Gopakumar-Vafa invariants of diagonal local $\mathbb{P}^1\times\mathbb{P}^1$.
One can see that $J_\text{pert}(\mu)$ and $J_\text{WS}(\mu)$
agree with the correct grand potential of ABJM theory.
However, $J_\text{np}(\mu)$ in \eqref{JBorel}
is different from the true membrane instantons in ABJM theory.
From the numerical fitting, we find that the leading term in 
$J_\text{np}(\mu)$ is simply given by
\begin{align}
 J_\text{np}^\text{(1)}(\mu)=\frac{1}{\pi\sin\frac{\pi k}{2}}\left(2\mu+1+\frac{\pi k}{2}\cot\frac{\pi k}{2}\right)\re^{-2\mu}.
 \label{BorelM2}
\end{align}
Note that this is the same, up to an overall numerical factor, 
as the non-perturbative corrections in the resolved conifold 
\eqref{eq:F-coni-np} with the identification $2\mu=2\pi T/g_s$ and $k=4\pi/g_s$.

\begin{figure}[tb]
\begin{center}
\begin{tabular}{ccc}
%\hspace{-3mm}
\resizebox{45mm}{!}{\includegraphics{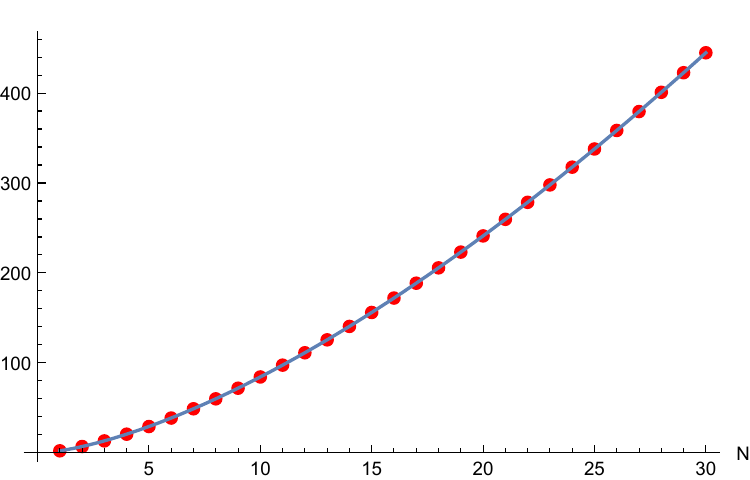}}
\hspace{2mm}
&
\resizebox{45mm}{!}{\includegraphics{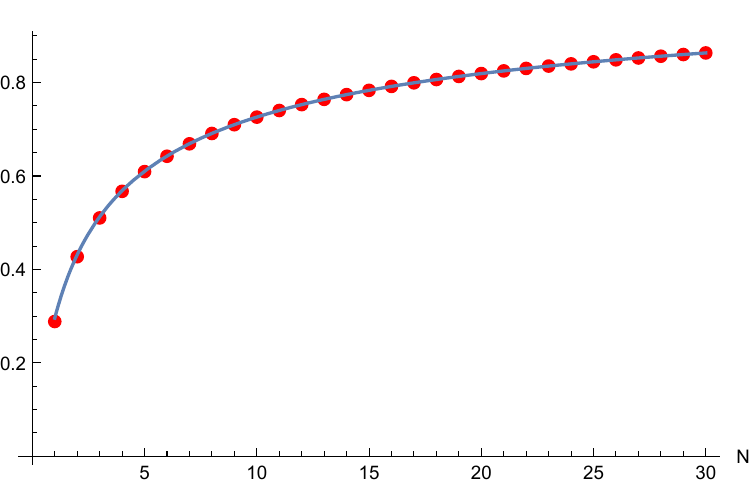}}
\hspace{2mm}
&
\resizebox{45mm}{!}{\includegraphics{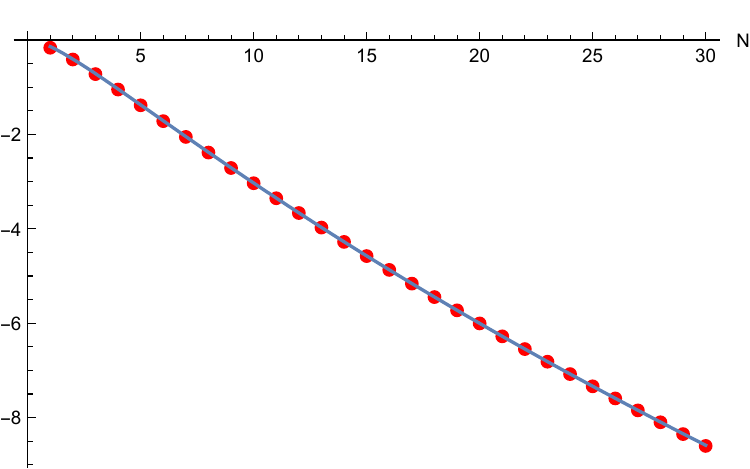}}\\
(a) Perturbative part &
(b) Worldsheet 1-instanton &
(c) Membrane-like correction
\end{tabular}
\end{center}
  \caption{We show the plot of resummed free energy $\mathcal{S}_0F$ of ABJM theory
 as a function of $N$ for $k=17/5$.
The solid curves represent some combinations of Airy functions computed from
$J_\text{Borel}(\mu)$ while the red dots are the
numerical values of $\mathcal{S}_0F$.
More precisely, each figure represent:
(a) the (minus of) free energy $-\mathcal{S}_0F$ and 
the perturbative free energy $-\log Z_\text{pert}$.
(b) The worldsheet 1-instanton normalized by the exponential factor
$\re^{2\pi\rt{2\la}}$: $(\re^{\mathcal{S}_0F}/Z_\text{pert}-1)\re^{2\pi\rt{2\la}}$
and $Z_\text{WS}^{(1)}\re^{2\pi\rt{2\la}}$. (c)
The membrane-like correction  normalized by the exponential factor
$\re^{k\pi\rt{2\la}}$: $(\re^{\mathcal{S}_0F}/Z_\text{pert}-1-Z_\text{WS}^{(1)})\re^{k\pi\rt{2\la}}$
and $Z_\text{np}^{(1)}\re^{k\pi\rt{2\la}}$.
}
  \label{fig:ABJM-resum}
\end{figure}

Our method of numerical fitting is as follows.
To compare  the Borel resummation $\mathcal{S}_0F$
and \eqref{BorelM2},
we first expand the right hand side of \eqref{Jborelint}  as in \cite{HMO2}
\begin{align}
 \int_C\frac{\rd \mu}{2\pi\ri}\,\re^{J_\text{Borel}(\mu)-N\mu}=Z_\text{pert}(N,k)
\Big[1+Z_\text{inst}(N,k)\Big],
\label{JboreltoZ}
\end{align} 
where the perturbative part is given by the Airy function
\begin{align}
 Z_\text{pert}(N,k)=C(k)^{-\frac{1}{3}}\re^{A_\text{c}(k)}\text{Ai}\Big[C(k)^{-\frac{1}{3}}(N-B(k))\Big].
\end{align}
The ``instanton'' corrections $Z_\text{inst}(N,k)$ in \eqref{JboreltoZ}
are also written as some combination of Airy functions.
In figure \ref{fig:ABJM-resum}, we show the plot of
the Borel resummation $\mathcal{S}_0F$ of ABJM theory
as a function of $N$ for $k=17/5$.
As we can see in figure \ref{fig:ABJM-resum}(c), the Borel resummation $\mathcal{S}_0F$
indeed contains the membrane-like contribution 
coming from \eqref{BorelM2}.
In this way, we have checked that \eqref{BorelM2}
reproduces the membrane-like corrections in $\mathcal{S}_0F$
for various other values of $k$.

On the other hand, the correct membrane 1-instanton
term in ABJM grand potential has the form
\begin{align}
 J_\text{M2}^{(1)}(\mu)=\big[a_1(k)\mu^2+b_1(k)\mu+c_1(k)\big]\re^{-2\mu}.
\end{align}
The analytic forms of the coefficients $a_1(k)$, $b_1(k)$ and $c_1(k)$ were determined
in \cite{HMO2}.
For example, the coefficient $b_1(k)$ of $\mu \re^{-2\mu}$ term in ABJM theory is given by
\begin{align}
 b_1(k)=\frac{2\cos^2\frac{\pi k}{2}}{\pi\sin\frac{\pi k}{2}},
\end{align}
which is clearly different from the coefficient of $\mu\re^{-2\mu}$ in \eqref{BorelM2}.
Also, there is no $\mu^2\re^{-2\mu}$ term in \eqref{BorelM2}.

It is interesting to observe that $J_\text{np}^{(1)}(\mu)$
has a correct pole structure at even integer $k$,
which precisely cancel the divergence coming from 
the worldsheet instantons
 \begin{align}
  \lim_{k\to2}\left[J_\text{WS}^{(1)}(\mu)+J_\text{np}^{(1)}(\mu)\right]&=
 \left[\frac{2\mu^2+2\mu+1}{\pi^2}+\frac{1}{2}\right]\re^{-2\mu},\nn
 \lim_{k\to4}\left[J_\text{WS}^{(2)}(\mu)+J_\text{np}^{(1)}(\mu)\right]&=
  -\left[\frac{2\mu^2+2\mu+1}{2\pi^2}+\frac{3}{2}\right]\re^{-2\mu},\nn
  \lim_{k\to6}\left[J_\text{WS}^{(3)}(\mu)+J_\text{np}^{(1)}(\mu)\right]&=
 \left[\frac{2\mu^2+2\mu+1}{3\pi^2}+\frac{83}{18}\right]\re^{-2\mu}.
 \end{align}
 This is consistent with the observation \cite{GMZ}
 that the Borel resummation of ABJM free energy is always finite for all $k\in\mathbb{R}$,
even at integer (or rational) values of $k$.
Since the poles are also canceled in ABJM theory,
this implies that the difference $J_\text{M2}(\mu)-J_\text{np}(\mu)$ is a regular function of $k$.

We can perform the resummation directly in the worldsheet instanton part \eqref{JGV} 
of the grand potential. In this approach,
 we also find the same non-perturbative correction in \eqref{BorelM2}.
Note that, in the genus expansion of the worldsheet instanton part,
the $(2g)!$ growth arises only from the $n_{g=0}^d$ terms in  \eqref{JGV}.
Therefore, it is natural to conjecture that
the non-perturbative correction in \eqref{JBorel}
comes only from the resummation of $n_{g=0}^d$ terms in  \eqref{JGV}.
If this is correct, this explains the similarity between 
\eqref{BorelM2} and the membrane instantons in the resolved conifold, since 
for the resolved conifold the 
only non-vanishing Gopakumar-Vafa invariants 
is $n_{g=0}^1=1$.

\section{Conclusions}
In this paper, we worked on a generalization of the Borel resummation.
An advantage of the generalized Borel resummation is that one can write down exact
integral representations.
Using such an integral representation, we showed that the resummation of the resolved
conifold free energy automatically contains the non-perturbative correction in $g_s$
in addition to the well-known Gopakumar-Vafa formula in the large radius limit $T \to \infty$.
Quite remarkably, this non-perturbative correction perfectly agrees with the earlier conjecture
for the membrane instanton correction to the topological string free energy in \cite{HMMO}.
We also gave various other examples, in which our resummation procedure works.
In particular, it is interesting that the resummation of the vortex contribution in the double
sine function contains anti-vortex contribution, which is non-perturbative
in the ellipsoid parameter $b$.

We also revisited the Borel resummation of the ABJM free energy,
and discussed quantitatively that the Borel resummation already has a membrane-like 
non-perturbative correction.
Interestingly, this correction precisely removes the poles that the worldsheet instanton correction has.
However, the membrane instanton-like correction in the Borel resummation 
do not reproduce the true membrane instanton correction in \cite{HMMO},
as has already been observed in \cite{GMZ}.
We need additional corrections to reproduce the exact result.
As noted in \cite{DMP2, GMZ}, these additional corrections should be explained by 
complex Borel singularities on the right half plane.
Up to now, there is no systematic way to treat these corrections quantitatively in the Borel analysis.
It would be important to tackle this problem.

Recently, a new non-perturbative formulation of the topological strings was proposed in \cite{GHM1},
based on the earlier work \cite{ACDKV}.
This formulation relates the topological string free energy to certain quantum spectral problem
(see also \cite{Huang:2014eha, Kashaev:2015kha, MZ, KMZ, Wang:2015wdy}).
It was conjectured in \cite{GHM1} that the exact spectral determinant in the spectral problem 
can be constructed by the non-perturbative free energy in \cite{HMMO}.
In the setup in \cite{HMMO}, the K\"ahler moduli of the Calabi-Yau geometry are
shifted by the B-field effect, as in \eqref{eq:t-B}.
As in the resolved conifold studied in this paper, this B-field shift perhaps makes
the genus expansion Borel summable.
If one considers the situation without the B-fields,
the genus expansion might be non-Borel summable.
As seen in subsection~\ref{sec:non-Borel}, in this case, it is natural to consider the trans-series expansion.
It would be very interesting to explore non-perturbative corrections in the trans-series.
The resurgence approach along the line \cite{MSW, Marino2008, SESV, CSESV} 
will be useful for this purpose.
Currently, a relationship between the non-perturbative corrections in the resurgent trans-series 
and those in the M-theoretic expansion is unclear.
It would be interesting to clarify it in detail.

\acknowledgments{We thank Daniele Dorigoni, Masazumi Honda and Marcos Mari\~no for various discussions.
We are grateful to Marcos Mari\~no for conceptually significant comments on the manuscript.}

\end{document}